\documentclass[12pt,english]{article}
\usepackage[breaklinks,hypertexnames=false]{hyperref}
\hypersetup{colorlinks=true,linkcolor=blue,urlcolor=blue,citecolor=blue}
\usepackage{amsthm}
\usepackage{amsmath}
\usepackage{amssymb}
\usepackage{esint}
\usepackage[authoryear]{natbib}

\usepackage{subcaption} % For subfigures

\usepackage{appendix}
\usepackage[english]{babel}
\usepackage{enumitem}
\usepackage{booktabs}
\usepackage{multirow}
\usepackage{natbib}
\usepackage{textcomp,mathcomp}
\usepackage{adjustbox}
\usepackage{dcolumn}
\usepackage{booktabs}
\usepackage{mathabx} % for widebar

\makeatletter
\usepackage{graphicx}
\usepackage{lscape}
\usepackage{rotating}
\usepackage[english]{babel}
\usepackage{latexsym}	
\usepackage{bbm}
\usepackage{tikz}
\usetikzlibrary{arrows,shapes,trees,positioning}
\usepackage{seqsplit}

\hypersetup{
	colorlinks=true,
	citecolor=blue
}

\makeatother

\newtheorem{definition}{Definition}

\newcommand{\defeq}{\vcentcolon=}

\input{tcilatex}
\usepackage{babel}

\usepackage{tikz}
\usetikzlibrary{arrows.meta, positioning, calc} % <-- Added 'calc'

\begin{document}

	\title{{{\huge Back to Feedback}\\{\Large Dynamics and Heterogeneity in Panel Data}}\thanks{I thank {two anonymous referees,} Jaap Abbring, Manuel Arellano, Dmitry Arkhangelsky, Irene Botosaru, Cl\'ement de Chaisemartin, Kevin Dano, Xavier D'Haultfoeuille, Dalia Ghanem, Bryan Graham, {Michael Knaus}, Thibaut Lamadon, Wooyong Lee, Chris Muris, {Mikkel Plagborg-M\o ller, Zach Shahn}, Alex Torgovitsky, and Kaspar Wuthrich for detailed comments and helpful discussions. This article is based on a presentation given at the Sargan Lecture of the Royal Economic Society in Birmingham in July 2025.}}

\author{St\'{e}phane Bonhomme \\ University of Chicago\\  }
\date{\today}
\maketitle
\vskip 1cm

\begin{abstract}
	
Many popular estimation methods in panel data rely on the assumption that the covariates of interest are strictly exogenous. However, this assumption is empirically restrictive in a wide range of settings. In this article I argue that credible empirical work requires meaningfully relaxing strict exogeneity assumptions. Econometricians have developed methods that allow for sequential exogeneity, which in contrast with strict exogeneity allows for the presence of feedback from past outcomes to future covariates or treatments. I review some of the classic work on linear models with constant coefficients, and then describe some approaches that allow for coefficient heterogeneity in models with feedback. Finally, in the last two parts of the article I review recent work that allows for sequential exogeneity in nonlinear panel data models, and mention possible extensions to network settings.

	\bigskip
	
	
\end{abstract}

\clearpage

\global\long\def\ind{\mathbb{1}}
\global\long\def\d{\mathrm{d}}
\global\long\def\t{\intercal}
\global\long\def\RR{\mathbb{R}}
\global\long\def\defeq{:=}

\section{Introduction}

A central concern during the past thirty years in applied (micro) economics has been the presence of un-modeled heterogeneity. Some of the most influential methodological advances, such as work on local average treatment effects (\citealp{imbens1994estimation}) and difference-in-differences (\citealp{abadie2005semiparametric}, \citealp{goodman2021difference}) have been motivated by the desire to allow for rich individual heterogeneity in treatment responses.   

However, if one looks back to the pre-``empirical credibility revolution'' period, a chief concern of econometricians then was to correctly specify dynamic processes of outcome and covariates. David Hendry's textbook summarizes many of the advances made during that previous era (\citealp{hendry1995dynamic}). As it stands, the focus on unobserved heterogeneity appears to have shifted attention away from earlier concerns about dynamics.

This situation is unfortunate. Presumably, both dynamics and heterogeneity are present in many of the empirical settings that applied economists study. Overlooking one misspecification concern to solely focus on another one hardly seems to be a recipe for credible research. Yet, today's applied work, though often based on methods that are mindful of heterogeneity, overwhelmingly relies on restrictive assumptions about un-modeled dynamics of covariates and outcomes.   

A key concept when introducing dynamics is the \emph{feedback process} -- the way past outcomes and covariates influence (i.e., ``feed back'' into) later covariates. When a covariate is a treatment of interest, this process is a dynamic analog to the propensity score, which depends on past values of outcome and treatment. In panel data applications, the feedback process may additionally vary between individuals. 

Allowing for feedback appears \emph{a priori} reasonable: after all, why should a covariate today (for example, the value of the minimum wage) not depend on past outcomes (such as employment or earnings)? Yet, a large body of empirical work relies on an assumption that rules out feedback entirely. This assumption, known as \emph{strict exogeneity} (SE), is at the heart of the textbook  justification for two-way fixed effects estimators, and it is either explicitly or implicitly assumed in most applications of difference-in-differences.   

Several arguments are commonly mentioned to support an assumption of strict exogeneity. One argument is based on the use of pre-trend checks. Another argument is that feedback is less likely when the treatment is aggregate (such as a policy reform) and outcomes are individual-specific. I will argue that {neither argument is} fully compelling, and that feedback bias is likely present in many difference-in-differences settings. Rather than dismissing outright the possibility that the treatment may in fact respond to past outcomes, it seems reasonable to study how one can address the presence of feedback empirically.

The goal of this article is to highlight existing research on un-modeled dynamics and feedback. The working assumption is that of \emph{sequential exogeneity} (SeqE). Under sequential exogeneity, the errors in the regression are assumed to be uncorrelated with past and current covariates, however they are allowed to correlate with future covariates. Thus, while SeqE may still be restrictive since it rules out contemporaneous selection, it is less restrictive than SE since it allows future treatments to be chosen based on past outcomes. This is a crucial, yet often under-appreciated, difference.

After introducing the main concepts and describing their implications for common estimators, I survey some ``classic'' approaches in the literature on dynamic panels. This includes the popular estimators of \citet{arellano1991some} and \citet{blundell1998initial}; see \citet{arellano2003panel} for a textbook treatment. However, the methods developed in this literature suffer from two main drawbacks. A first challenge is the proliferation of instruments in dynamic settings and the resulting instability of estimators. I review several approaches that have been proposed to improve over the original dynamic panel methods. A second challenge is that classic dynamic panel methods are vulnerable to the presence of un-modeled heterogeneity. Although heterogeneity in the model is limited to the intercept (the so-called ``fixed effect'', which is differenced out), heterogeneity in coefficients -- reflecting treatment effects heterogeneity -- may be empirically prevalent, as emphasized by \citet{marx2024heterogeneous}. 

An important goal of the article is to cover some of the ``modern'' approaches to models with feedback, including approaches that allow for coefficient heterogeneity. The starting point is a seminal paper by Gary Chamberlain (\citealp{chamberlain2022feedback}), which was first circulated in the 1990s and remains highly relevant to today's methodological and empirical work. This paper establishes a negative result, showing that, in a simple model with a binary sequentially exogenous covariate and two periods, averages of coefficients (i.e., average treatment effects on subpopulations) are not identified. This stands in contrast with the strictly exogenous case, where average effects for individuals whose covariate changes over time (the so-called ``movers'') are identified (e.g., \citealp{chamberlain1992efficiency}).   

There are two possible reactions to Chamberlain's negative result. The first one is to focus on quantities that are identified, as in \citet{bonhomme2025unrestricted}. The second one is to relax the identification requirement and bound the estimand of interest, as exemplified by \citet{lee2026identification}. Since addressing dynamic aspects is in my view key for credible inference, more work is needed in this area, with the aim to account for heterogeneity similarly to the ``new difference-in-differences literature'' (e.g., \citealp{de2023two}, \citealp{sun2021estimating}, \citealp{callaway2021difference}, \citealp{borusyak2024revisiting}) while relaxing the often implausible assumption of strict exogeneity. 

Another ``modern'' question in the literature is to relax the assumption that the model is linear in parameters. Deriving valid moment restrictions on parameters is challenging in nonlinear settings, due to the twin presence of heterogeneity and feedback. Solutions exist in specific models, such as Poisson regressions and multiplicative proportional hazard models. I also review recent work by \citet{bonhomme2025feedback} that extends the functional differencing approach introduced in \citet{bonhomme2012functional} to models with feedback. Following \citet{bonhomme2023identification}, I further explain how the identified set of parameters can be characterized using linear programming, generalizing the analysis in \citet{honore2006bounds} to models with feedback.

Lastly, an important avenue is the study of dynamic network settings. Popular estimators such as the AKM estimators of \citet{abowd1999high} have been used in a variety of settings to understand the sources of wage dispersion, among many other questions. However, the validity of this approach hinges on strict exogeneity. Allowing for feedback in network contexts may be as or more relevant than in single-agent panel data contexts. For example, the assumption in AKM that job mobility does not depend on past wages is controversial. Although the literature on the topic is still in its infancy, I conclude the article by mentioning some recent research and possible approaches.

\section{Sequential exogeneity and feedback}

\subsection{Definitions in linear models}

Strict and sequential exogeneity are central concepts in panel data. To introduce them, it is useful to start with a linear model for a single time series (that is, focusing on a single individual in the panel). Specifically, consider a linear time-series model of the form
\begin{equation}
	Y_{t}=X_{t}'\beta+U_{t},\quad t=1,...,T,\label{mod_TS}
\end{equation}
where $Y_t$ is a scalar outcome and $X_t$ is a vector of covariates. 

\begin{definition}{(strict and sequential exogeneity, linear models)}$\quad$\label{def1}
	
	\noindent $X_t$ is strictly exogenous (SE) if \begin{equation}\mathbb{E}[U_t\,|\, X_1,...,X_T]=0, \quad t=1,...,T.\end{equation}
	
	\noindent $X_t$ is sequentially exogenous (SeqE) if  \begin{equation}\mathbb{E}[U_t\,|\, X_1,...,X_t]=0, \quad t=1,...,T.\end{equation}

\end{definition}

The two types of exogeneity introduced in Definition \ref{def1} are classical concepts in time series analysis, and they can be found under various names in the literature. Strict exogeneity is often referred to as ``strong exogeneity'', and contrasted with ``weak'' (i.e., sequential) exogeneity; see for example \citet{engle1983exogeneity}. Another common term for sequential exogeneity is ``predeterminedness''.

Strict exogeneity requires covariates $X_s$ in \emph{all} periods $s=1,...,T$ to be unrelated to the outcome disturbances $U_t$. In contrast, sequential exogeneity only requires covariates $X_s$ in the \emph{past and current} periods $s=1,...,t$ to be unrelated to $U_t$. The key difference between the two assumptions is that, unlike SE, SeqE allows for \emph{feedback}: outcome realizations associated with some shocks $U_t$ are allowed to influence future realizations of the covariates $X_{t+1},...,X_T$.

\begin{figure}[tbp]
	{	\centering
		\caption{An illustration of strict and sequential exogeneity\label{fig_strict_exog}}
		
		\begin{minipage}[t]{0.5\textwidth}
			\centering
			A. Strict exogeneity (SE)\\[4pt]
			\begin{tikzpicture}[node distance=2cm, every node/.style={font=\large}]
				% Y nodes (top row)
				\node (ytm1) {$Y_{t-1}$};
				\node[right=of ytm1] (yt) {$Y_t$};
				\node[right=of yt] (ytp1) {$Y_{t+1}$};
				
				% X nodes (bottom row, offset left)
				\node[below left=1.5cm and 0.5cm of ytm1] (xtm1) {$X_{t-1}$};
				\node[right=of xtm1] (xt) {$X_t$};
				\node[right=of xt] (xtp1) {$X_{t+1}$};
				
				% Y arrows (solid)
				\draw[->, thick] (ytm1) -- (yt);
				\draw[->, thick] (yt) -- (ytp1);
				
				% X arrows (dashed)
				\draw[->, dashed, thick] (xtm1) -- (xt);
				\draw[->, dashed, thick] (xt) -- (xtp1);
				
				% Arrows from X to Y (solid)
				\draw[->, thick] (xtm1) -- (ytm1);
				\draw[->, thick] (xt) -- (yt);
				\draw[->, thick] (xtp1) -- (ytp1);
			\end{tikzpicture}
		\end{minipage}%
		\hfill
		\begin{minipage}[t]{0.5\textwidth}
			\centering
			B. Sequential exogeneity (SeqE)\\[4pt]
			\begin{tikzpicture}[node distance=2cm, every node/.style={font=\large}]
				% Y nodes (top row)
				\node (ytm1) {$Y_{t-1}$};
				\node[right=of ytm1] (yt) {$Y_t$};
				\node[right=of yt] (ytp1) {$Y_{t+1}$};
				
				% X nodes (bottom row)
				\node[below left=1.5cm and 0.5cm of ytm1] (xtm1) {$X_{t-1}$};
				\node[right=of xtm1] (xt) {$X_t$};
				\node[right=of xt] (xtp1) {$X_{t+1}$};
				
				% Y arrows (solid)
				\draw[->, thick] (ytm1) -- (yt);
				\draw[->, thick] (yt) -- (ytp1);
				
				% X arrows (dashed)
				\draw[->, dashed, thick] (xtm1) -- (xt);
				\draw[->, dashed, thick] (xt) -- (xtp1);
				
				% Arrows from X to Y (solid)
				\draw[->, thick] (xtm1) -- (ytm1);
				\draw[->, thick] (xt) -- (yt);
				\draw[->, thick] (xtp1) -- (ytp1);
				
				% New dashed arrows from Y to X
				\draw[->, dashed, thick] (ytm1) to[bend left=20] (xt);
				\draw[->, dashed, thick] (ytm1) to[bend left=25] (xtp1);
				\draw[->, dashed, thick] (yt) to[bend left=15] (xtp1);
			\end{tikzpicture}
		\end{minipage}
	}
	
	{\footnotesize{\textit{Notes: Schematic description of strict exogeneity (left panel) and sequential exogeneity (right panel), based on Definition \ref{def1}.}}}
	
\end{figure}
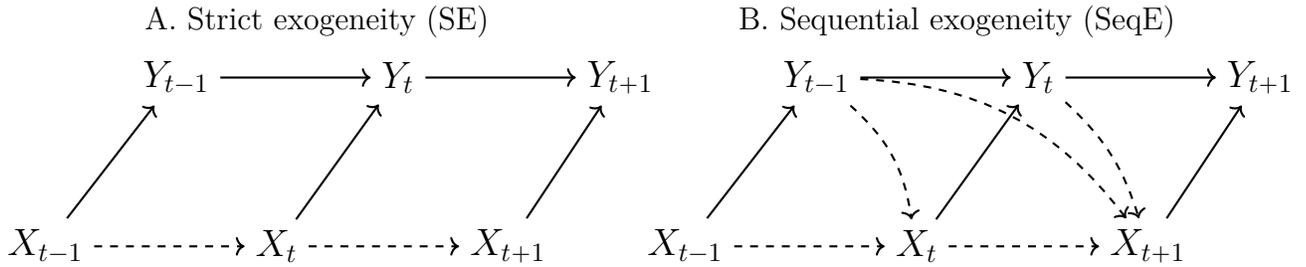

Consider next the case where the model includes a lagged outcome,\footnote{As a convention, I will assume throughout that $Y_0$ is part of the observation sample.} 
\begin{equation}
	Y_{t}=X_{t}'\beta+\rho Y_{t-1}+U_{t},\quad t=1,...,T.\label{mod_lag}
\end{equation}
In this dynamic model, strict and sequential exogeneity of $X_t$ are typically assumed to hold conditional on the history of past outcomes. Hence, SE becomes 
\begin{equation}\mathbb{E}[U_t\,|\, Y_{0},...,Y_{t-1},X_1,...,X_T]=0, \quad t=1,...,T,\label{eq_lag1}\end{equation}  
while SeqE becomes
\begin{equation}\mathbb{E}[U_t\,|\, Y_{0},...,Y_{t-1},X_1,...,X_t]=0, \quad t=1,...,T.\label{eq_lag2}\end{equation}  
Note that (\ref{eq_lag1}) and (\ref{eq_lag2}) imply that $U_t$ are serially uncorrelated.

Figure \ref{fig_strict_exog} represents the strict and sequential exogeneity assumptions graphically. In the left panel, the solid arrows indicate that $X_t$ has a direct effect on $Y_t$. In addition, the lagged outcome $Y_{t-1}$ may also directly affect $Y_t$, for example if the model includes a lag as in (\ref{mod_lag}). The graph in the left panel also features dashed arrows, i.e., possible relationships between covariates $X_t$ over time. However, it does not feature any link between past outcomes, such as $Y_{t-1}$, and current covariates $X_t$. In contrast, the right panel in Figure \ref{fig_strict_exog}, which represents the sequential exogeneity assumption, does allow for relationships from past outcomes $Y_{t-1}$ to current covariates $X_t$, indicated by dashed arrows.

\subsection{Definitions in nonlinear models with arbitrary heterogeneity}

These definitions can be generalized to models that are nonlinear and feature arbitrary heterogeneity. Doing so is useful given the widespread use of strict exogeneity assumptions for causal inference in difference-in-differences settings. 

{Following \citet{abbring2007econometric} -- and \citet{rambachan2019common} in a time-series setting -- let $Y_t(x_1,...,x_T)$ denote the potential outcome for given values of the covariates $X_1=x_1,...,X_T=x_T$. Suppose that the following \emph{no-anticipation} assumption is satisfied for all $t=1,...,T$:
\begin{equation}
	Y_t(x_1,...,x_t,x_{t+1},...,x_T)=Y_t(x_1,...,x_t,x_{t+1}',...,x_T'),\text{ for all }x_1,...,x_T,x_{t+1}',...,x_T'.\label{eq_NA}
\end{equation}
Assumption (\ref{eq_NA}) implies that the outcome is not causally affected by future treatments. While no-anticipation may be restrictive, it is commonly made in the literature and I will adopt it throughout this article. Under (\ref{eq_NA}) we may thus write $Y_t(x_1,...,x_t)$.\footnote{One could explicitly introduce lagged outcomes $Y_{t-s}$ using the potential outcome notation {$Y_t(x_1,...,x_t,y_0,y_1,...,y_{t-1})$} so as to model state dependence, as in \citet{torgovitsky2019nonparametric}.} Feedback
	and anticipation both generate statistical dependence between outcomes and later treatments, and the approach taken here is to allow for the former while ruling out the latter.

	%, so
	%allowing for the former requires ruling out the latter. 

In this general setup, one can write sequential exogeneity as the following conditional independence assumption:
\begin{equation}
	Y_{t+s}(x_1,...,x_{t+s}) \text{ for all }s\geq 0\text{ are independent of }X_{t} \text{ given }Y_0,...,Y_{t-1},X_1=x_1,...,X_{t-1}=x_{t-1}.\label{eq_SeqE}\end{equation}
That is, the treatment is assumed to be independent of future potential outcomes conditional on the history of treatment and past outcomes. In this form, Condition (\ref{eq_SeqE}) is sometimes called \emph{sequential exchangeability} (\citealp{robins1986new}, \citealp{marx2024heterogeneous}) or \emph{sequential randomization} (\citealp{bojinov2021panel}). {For example, writing
	$$Y_t(x_1,...,x_t)=m_t\left(x_1,...,x_t,U_t\right),$$
	where $U_t$ collects the shocks affecting $Y_t$, Condition (\ref{eq_SeqE}) requires that
	$X_t$ be independent of current and future shocks $U_{t+s}$, $s\geq 0$, conditional on past
	outcomes and treatments.}

Condition (\ref{eq_SeqE}) is helpful to understand the economic content of sequential exogeneity. Under (\ref{eq_SeqE}), the agent chooses $X_t$ given her information set, which includes her past treatment and outcome realizations (as well as heterogeneity and possibly other state variables included in $X_t$), but does not include current or future potential outcomes. This precludes advance information about returns, for example. 

In contrast, in this setup one can write strict exogeneity as
\begin{equation}	Y_{t'}(x_1,...,x_{t'}) \text{ for all }t'=1,...,T\text{ are independent of }X_{t} \text{ given }X_1=x_1,...,X_{t-1}=x_{t-1},\label{eq_SE}\end{equation}
where the information set does not include \emph{any} of the potential outcomes: past, present or future. This precludes agents responding to \emph{past} outcomes when choosing $X_t$, and requires the choice problem to be completely unrelated to outcome realizations.

As \citet{abbring2007econometric} argue, in combination with no-anticipation (\ref{eq_NA}), sequential exogeneity (\ref{eq_SeqE}) imposes strong restrictions on the information available to agents -- even though these are meaningfully weaker than the restrictions imposed by strict exogeneity (\ref{eq_SE}). Assuming that (\ref{eq_SeqE}) holds for all individuals implicitly requires that agents not act on information predictive of future outcomes beyond past covariates and outcomes. In panel data settings, it is important to allow in addition for an individual effect $A$, unobserved to the econometrician and  typically assumed to be invariant over time, such that (\ref{eq_SeqE}) holds conditionally on $A$ but not unconditionally. This conditional version of sequential exogeneity is the standard one in the panel data literature, and it is the one I adopt throughout this article.

Even when assumed to hold conditional on an individual effect $A$, sequential exogeneity remains potentially restrictive. Importantly, sequential exogeneity cannot accommodate the presence of simultaneity (when $X_t$ and $Y_t$ are simultaneously determined) or serially-correlated time-varying {confounders} (e.g., some unobserved covariate $V_t$ in (\ref{mod_TS}) that is correlated with $X_t$). In both these situations, instruments $Z_t$ external to the model would be required for identification. Nonetheless, the appeal of sequential exogeneity is that it captures any empirically plausible mechanisms through which past disturbances $U_{t-1},U_{t-2},...$ may correlate with current covariates $X_t$. }

In (linear or nonlinear) models with sequential exogeneity, the \emph{feedback process} is defined as follows.
\begin{definition}
	The feedback process is the conditional density of 
	$$X_t\,|\, Y_0,...,Y_{t-1},X_{1},...,X_{t-1}.$$
\end{definition}

\noindent In words, the feedback process is the conditional density of current covariates given past outcomes and covariates. In panel data, the feedback process is defined conditionally on the individual effect $A$. Under SE, the feedback process is simply the conditional density of the current covariate given the history of covariates (and $A$). From a time-series perspective, strict exogeneity thus rules out Granger causality from past $Y_s$, for $s=1,...,t-1$, to future $X_t$.\footnote{	\citet{chamberlain1982general} shows that the classic definition of strict exogeneity through conditional independence, 
	\begin{equation*}Y_t \text{ is independent of }X_{t+1},...,X_T \text{ conditional on }X_1,...,X_t,\end{equation*}
is equivalent to the following \emph{no feedback} condition:
	\begin{equation*}X_t \text{ is independent of }Y_0,...,Y_{t-1} \text{ conditional on }X_1,...,X_{t-1}.\end{equation*}}

\subsection{The perils of strict exogeneity}
Under strict exogeneity, the feedback process is conditionally independent of past outcomes. That is, SE rules out the presence of feedback. However, assuming no feedback is restrictive in many economic settings. 

Consider a model of job training, as described in \citet{arellano2003panel}, Chapter 8. Focusing on a single worker, let the earnings of the worker be $Y_t=Y_t^*+\beta X_t$, for $Y_t^*$ the earnings in the absence of training, {and $X_t$ a binary training indicator.} In this setting, strict exogeneity is likely to be violated since there is evidence of a dip in the earnings of training participants before training happens (\citealp{ashenfelter1985using}). One way to account for this dip is to model earnings in the absence of training $Y_t^*$ as a dynamic (e.g., autoregressive) process while allowing for {pre-training earnings $Y_{t-s}^*$, $s\geq 1$, to affect} {the training indicator $X_t$.} This type of approach based on sequential exogeneity will be the topic of Section \ref{sec_AB} onward.

In fact, feedback is a central mechanism in most dynamic economic models. Consider as an example a consumption model where $Y_t$ denotes household consumption and $X_t$ include household assets; see \citet{deaton1992understanding} for a comprehensive treatment. Through the budget constraint, assets today depend on how much the household consumed last period, which mechanically implies the presence of feedback and violates strict exogeneity. More generally, feedback is central to dynamic economic models with forward-looking agents which are popular in {industrial organization} (\citealp{rust1987optimal}), labor economics (\citealp{keane1997career}), and many other fields.

Yet, while feedback is economically plausible in many settings, strict exogeneity is routinely assumed in applications. A case in point is the popularity of difference-in-differences methods, which explicitly or implicitly rely on the assumption that the treatment of interest is strictly exogenous {(\citealp{chabe2015analysis}, \citealp{de2022not}, \citealp{ghanem2022selection}, \citealp{marx2024parallel}, \citealp{knaus2026causal}).} The next section will describe how strict exogeneity, when assumed in a setting where covariates are only sequentially exogenous, can create important biases in estimation.  
 
% Augmented with a lagged outcome and other covariates
% \begin{equation}
% 	Y_{t}=X_{t}'\beta+\rho Y_{t-1}+W_t'\gamma+U_{t},\quad t=1,...,T.
% \end{equation}
% 

\subsection{Fixed effects}

Panel data has become the leading data format used in empirical micro-economics. A common specification accounts for a latent individual-specific intercept (a ``fixed effect''), as in the model
\begin{equation}
Y_{it}=X_{it}'\beta+A_i+U_{it},\quad i=1,...,N,\quad  t=1,...,T.\label{eq_pan}
\end{equation}
In applications, the model is often augmented with a lagged outcome and other covariates, as in
\begin{equation}
Y_{it}=X_{it}'\beta+\rho Y_{i,t-1}+W_{it}'\gamma+A_i+U_{it},\quad i=1,...,N,\quad  t=1,...,T,\label{eq_dyn}
\end{equation}
where $W_{it}$ may include time effects and other strictly exogenous or sequentially exogenous covariates.

There are two common versions of sequential exogeneity in this context (abstracting from lagged outcomes $Y_{i,t-1}$ and additional covariates $W_{it}$ for simplicity). The first one is not conditional on the latent effect $A_i$:  
\begin{equation}\mathbb{E}[U_{it}\,|\, X_{i1},...,X_{it}]=0,\label{seqpan1}\end{equation}
while the second one is conditional on the latent effect $A_i$:
\begin{equation}\mathbb{E}[U_{it}\,|\, X_{i1},...,X_{it},A_i]=0.\label{seqpan2}\end{equation}
One can similarly define unconditional and conditional versions of strict exogeneity.\footnote{That is,
$\mathbb{E}[U_{it}\,|\, X_{i1},...,X_{iT}]=0$ (unconditional) and
$\mathbb{E}[U_{it}\,|\, X_{i1},...,X_{iT},A_i]=0$ (conditional).} In addition, in dynamic models such as (\ref{eq_dyn}), strict and sequential exogeneity are typically assumed to hold conditional on the history of past outcomes.
 
These notions can be extended to models with coefficient heterogeneity, such as 
\begin{equation}
	Y_{it}=X_{it}'B_i+C_i+U_{it},\quad i=1,...,N,\quad  t=1,...,T.\label{mode_hetero}
\end{equation}
In model (\ref{mode_hetero}), the unconditional and conditional versions of SeqE and SE are defined exactly as above, with $A_i=(B_i',C_i)'$. Section \ref{sec_hetero} will consider such extensions allowing for coefficient heterogeneity. This will reveal that, in models with coefficient heterogeneity, the unconditional and conditional versions of SeqE (\ref{seqpan1}) and (\ref{seqpan2}) have -- perhaps unexpectedly -- different implications for (partial) identification of average effects.

In panel data models such as (\ref{eq_pan}), it is common to leave the feedback process
	$$X_{it}\,|\, Y_{i0},...,Y_{i,t-1},X_{i1},...,X_{i,t-1},A_i$$
	unrestricted. This is conceptually appealing since this allows for arbitrary forms of feedback, based on the history of covariates and outcomes (which are often state variables in the economic model), in a way that accommodates arbitrary individual heterogeneity (for example, consistent with heterogeneous technology or preferences). 
	
	However, in some settings it may be appealing to impose some assumptions on the feedback process. \emph{Markovian feedback}, through which $X_{it}$ are independent of the history conditional on $Y_{i,t-1}$, {$X_{i,t-1}$}, and $A_i$, say, is commonly assumed in structural models. \emph{Homogeneous feedback}, through which $X_{it}$ are independent of $A_i$ given the history $Y_{i0},...,Y_{i,t-1},X_{i1},...,X_{i,t-1}$, is often assumed in sequential experiments (as in \citealp{robins1986new}) and may be plausible in some economic models, such as models with a homogeneous state transition equation. Section \ref{sec_nonlin} will return to the case of restricted feedback.

\section{Feedback bias}

An important statistical consequence of the presence of feedback is that the OLS estimator is biased for finite $T$ {(\citealp{Kendall1954}, \citealp{White1961Asymptotic}).} This bias affects popular fixed effects regressions in panel data, where the bias of the estimator in the time series translates into panel data inconsistency as the cross-sectional size of the sample $N$ grows while the number of time periods $T$ remains fixed. 

\subsection{Bias in time series}

Consider first the time-series model (\ref{mod_TS}). The OLS estimator is given by (assuming that $\sum_{t=1}^T X_tX_t'$ is non-singular):
$$\widehat{\beta}=\left(\sum_{t=1}^T X_tX_t'\right)^{-1}\sum_{t=1}^T X_tY_t.$$
The OLS estimator $\widehat{\beta}$ is unbiased under strict exogeneity, since in this case
$$\mathbb{E}[\widehat{\beta}]=\beta+\mathbb{E}\left[\left(\sum_{t=1}^T X_tX_t'\right)^{-1}\sum_{t=1}^T X_t\underset{=0}{\underbrace{\mathbb{E}\left[U_t\,|\, X_1,...,X_T\right]}}\right]=\beta.$$
However, $\widehat{\beta}$ is generally biased under sequential exogeneity.

To see the source of bias, consider the case where there are $T=2$ periods. We have
$$\widehat{\beta}=\beta+\underset{(I)}{\underbrace{(X_1X_1'+X_2X_2')^{-1}X_1U_1}}+\underset{(II)}{\underbrace{(X_1X_1'+X_2X_2')^{-1}X_2U_2}}.$$
Under SE, both terms (I) and (II) have zero mean. Under SeqE, (II) still has zero mean, since $\mathbb{E}[U_2\,|\, X_1,X_2]=0$, so
$$\mathbb{E}[(II)]=\mathbb{E}\left[(X_1X_1'+X_2X_2')^{-1}X_2\underset{=0}{\underbrace{\mathbb{E}[U_2\,|\, X_1,X_2]}}\right]=0.$$ 
However, $U_1$ may be correlated with $X_2$ under SeqE because of the presence of feedback. As a result, we generally have
$$\mathbb{E}[(I)]=\mathbb{E}\left[(X_1X_1'+X_2X_2')^{-1}X_1\underset{\neq 0}{\underbrace{\mathbb{E}[U_1\,|\, X_1,X_2]}}\right]\neq 0.$$

In a time-series setting, if sequential exogeneity holds, then $\widehat{\beta}$ is consistent as $T$ tends to infinity under standard conditions, despite the presence of bias. A key property for consistency of OLS is that a lack of contemporaneous correlation property is satisfied, 
$$\mathbb{E}[X_tU_t]=0,\quad t=1,...,T.$$ 
See for example \citet{hayashi2011econometrics}, Chapter 2.\footnote{Although \citet{hayashi2011econometrics} refers to $\mathbb{E}[X_tU_t]=0$ as regressors being ``predetermined'', this terminology differs from what the panel data literature defines as ``predetermined'' (i.e., sequentially exogenous) regressors.}

However, the bias can be large for moderate $T$. Recently, \citet{mikusheva2023linear} study the properties of OLS in linear regression models with a large number of sequentially exogenous covariates relative to sample size. Focusing on an asymptotic regime where the number of regressors increases with $T$, they document that biases may be so large as to render the estimator inconsistent.

\subsection{Inconsistency in short panel data}

The bias of OLS for short $T$ is particularly problematic in empirical applications to short panel data. Consider the linear panel data model with fixed effects
\begin{equation}
	Y_{it}=X_{it}'\beta+A_i+U_{it},\quad i=1,...,N,\quad  t=1,...,T,\label{PD_model}
\end{equation}
under the assumption that the regressors $X_{it}$ are sequentially exogenous\footnote{Note that here the sequential exogeneity assumption is not conditional on the fixed effect $A_i$, but the conditional SeqE assumption (\ref{seqpan2}) has the same bias implications.
}
\begin{equation}\mathbb{E}[U_{it}\,|\, X_{i1},...,X_{it}]=0.\label{eq_seq}
\end{equation}

Let $\widehat{\beta}$ denote the OLS estimator in (\ref{PD_model}) -- with fixed effects, so $\widehat{\beta}$ coincides with the so-called ``within-group'' estimator. Typically, one can show that, as $N\rightarrow \infty$ for $T$ fixed we have
$$\underset{N\rightarrow \infty}{\limfunc{plim}} \, \widehat{\beta}=\beta+C_T\neq \beta,$$
where $C_T$ is a non-zero constant. 

To provide additional intuition on the bias, it is informative to expand $C_T$ as $T$ tends to infinity. Under suitable regularity conditions, we obtain the following expansion: 
\begin{equation}\underset{N\rightarrow \infty}{\limfunc{plim}} \, \widehat{\beta}=\beta+\frac{C}{T}+o\left(\frac{1}{T}\right),\label{bias_expand}\end{equation}
for some constant $C$. This shows that, to some approximation, the bias of $\widehat{\beta}$ decays as $T$ increases in a way that is inversely proportional to $T$. Hence, one expects that, at least when $T$ is not too small, the bias will be approximately divided by two when the number of time periods doubles. However, in short panels, the bias can be substantial.\footnote{The OLS estimator in first differences is similarly inconsistent for fixed $T$ in the presence of feedback. However, in contrast with the within-group estimator, it remains inconsistent as $T$ tends to infinity in general (e.g., \citealp{alvarez2003time}).} 

An important example is the so-called ``Nickell bias'' in an autoregressive panel data model where $X_{it}$ is scalar and coincides with the lagged outcome $Y_{i,t-1}$. \citet{nickell1981biases} derives an explicit expression for the probability limit of $\widehat{\beta}$ as $N$ tends to infinity for fixed $T$, and \citet{alvarez2003time} derive the asymptotic properties of $\widehat{\beta}$ as $N$ and $T$ tend to infinity jointly. Under assumptions that include i.i.d. homoskedastic errors and stationary initial conditions, \citet{alvarez2003time} show that
$$\underset{N\rightarrow \infty}{\limfunc{plim}} \, \widehat{\beta}=\beta-\frac{1+\beta}{T}+o\left(\frac{1}{T}\right).$$
More generally, the presence of feedback bias represents a major challenge in panel data models with
feedback and heterogeneity.

\section{Feedback bias and practice}

The presence of feedback bias has important implications for the practice of difference-in-differences, as this section and the next illustrate. 

%Dynamic bias from omitting lagged outcomes.

\subsection{Dynamic bias in difference-in-differences and event studies}

Consider the \emph{two-way fixed effects} model popular in the current applied literature
\begin{equation}
	Y_{it}=X_{it}\beta+A_i+F_t+U_{it},\quad i=1,...,N,\quad  t=1,...,T,\label{eq_TWFE_mod}
\end{equation}
in the case of a scalar $X_{it}$ and abstracting from other covariates for simplicity. Consider the OLS estimator with unit and time fixed effects, which coincides with the standard \emph{two-way fixed-effects} estimator (here for a balanced panel),
$$\widehat{\beta}^{\rm TWFE}= \frac{\sum_{i=1}^N\sum_{t=1}^T \overset{..}{X}_{it}\overset{..}{Y}_{it}}{\sum_{i=1}^N\sum_{t=1}^T\overset{..}{X}_{it}^2},$$
where $\overset{..}{Z}_{it}=Z_{it}-\frac{1}{T}\sum_{s=1}^TZ_{is}-\frac{1}{N}\sum_{j=1}^NZ_{jt}+\frac{1}{NT}\sum_{j=1}^N\sum_{s=1}^TZ_{js}$ denotes the double difference of $Z_{it}$. Under strict exogeneity, in model (\ref{eq_TWFE_mod}) with constant coefficients, $\widehat{\beta}^{\rm TWFE}$ is unbiased for $\beta$, and consistent in short panels under mild additional conditions. However, $\widehat{\beta}^{\rm TWFE}$ is generally \emph{inconsistent} in short panels when $X_{it}$ {is} sequentially exogenous. 

It is important to point out that the nature of the bias is different from the issues highlighted by the recent literature on difference-in-differences (e.g., \citealp{goodman2021difference}, \citealp{de2023two}, \citealp{sun2021estimating}, \citealp{callaway2021difference}). In that literature, the presence of un-modeled heterogeneity in coefficients affects the {interpretation of the probability limit} of popular estimators such as two-way fixed effects. In contrast, here the (homogeneous) coefficient of interest is biased due to the presence of un-modeled dynamics, through feedback. Presumably both forces may be simultaneously at play in the data, and Section \ref{sec_hetero} will study models with both feedback \emph{and} coefficient heterogeneity.

As \citet{ashenfelter1985using} point out, strict exogeneity may be restrictive in the context of estimating the effect of training programs using difference-in-differences methods. Participation in the program, indicated by $X_{it}=1$, may respond to realizations of past outcomes. In the presence of sequential exogeneity, the two-way fixed-effects estimator is generally inconsistent, with a bias $C_T$ inversely related to the panel length as in (\ref{bias_expand}). However, today's applied work often provides too little discussion of how the treatment was determined and whether it is affected by past outcomes through feedback, which are key to argue for the plausibility of strict exogeneity.

	The exact same issue affects event study regressions. Consider as an example	the model
	\begin{equation}
		Y_{it}=\sum_{s=-a}^bX_{i,t-s}\beta_s+A_i+F_t+U_{it},\quad i=1,...,N,\quad  t=1,...,T,
	\end{equation}
which controls for leads and lags of $X_{it}$. This is a popular model for policy evaluation. While the OLS estimators of $\beta_{-a},...,\beta_b$ (with unit and time fixed effects) are unbiased when $X_{it}$ is strictly exogenous, they are generally biased -- and inconsistent in short panels -- when $X_{it}$ is sequentially but not strictly exogenous. This shows that event study estimates, similarly to two-way fixed-effects estimates, crucially rely on strict exogeneity.

Finally, other popular extensions of two-way fixed-effects methods are not immune to feedback bias either. Interactive fixed-effects approaches are typically only consistent as both $N$ and $T$ tend to infinity (\citealp{bai2009panel}, \citealp{moon2017dynamic}), while consistency in short panels of fixed-$T$ versions such as \citet{ahn2013panel} requires strict exogeneity.\footnote{Consistency of synthetic control methods (\citealp{abadie2010synthetic}) and synthetic difference-in-differences (\citealp{arkhangelsky2021synthetic}) requires the number of pre-treatment periods to be large.}

\subsection{A cautionary note on pre-trend checks\label{subsec_preT}}

In difference-in-differences, researchers routinely run pre-trend checks to validate their empirical approaches. In some settings, such checks suggest clear violations of strict exogeneity. As a recent example, consider the analysis in \citet{acemoglu2019democracy}, who study how the spread of democracy during the period 1960-2010 has affected economic growth. Using a panel of countries, they
regress GDP per capita on a democracy indicator, lags of GDP per capita, and country and year fixed effects. The inclusion of a country effect accounts for permanent GDP differences between countries, and lags of GDP are included to make the error term serially uncorrelated. The authors document graphically that democratization episodes are, on average, preceded
by a temporary dip in GDP, and argue this shows a clear violation of the parallel trends assumption that underlies difference-in-differences and other panel data methods based on strict exogeneity.

%(see Figure \ref{fig:figgrowthpng}). 

%\begin{figure}[tbp]
%	\begin{center}
%		\caption{Dip in GDP per capita pre democratization\label{fig:figgrowthpng}}
%		\includegraphics[width=8cm]{fig_growth.png}
%	\end{center}
%	{\footnotesize{\textit{Notes: Figure from \citet{acemoglu2019democracy}.}}}
%\end{figure}

However, there is growing evidence against using pre-trend checks as a formal validation of strict exogeneity (or parallel trends) assumptions. \citet{roth2022pretest} points out two main issues with the current practice of pre-trend checks. First, pre-trend tests often have low power. Second, they are only an indirect test of model validity. The limitations of pre-trend checks have been noted by a number of authors, and have motivated some recent methodological developments in difference-in-differences settings (e.g., \citealp{freyaldenhoven2019pre}, \citealp{rambachan2023more}). 

Recently, \citet{ghanem2022selection} revisit the \citet{lalonde1986evaluating} analysis of the effect of job training on earnings.
They find that, while a pre-trend test does not reject strict exogeneity,
two-way fixed effects and other difference-in-differences estimators differ substantially from the experimental benchmark. This suggests that, while clear rejections such as suggested by the graphical evidence in \citet{acemoglu2019democracy} are informative, non-rejections of strict exogeneity should be interpreted with caution. We will return to the limitations of pre-trend checks in the next section, in the context of an example.

\subsection{Dynamic bias when treatments are aggregate}

A common argument against strict exogeneity assumptions is that, when covariates are choice variables, it is natural to expect that they may respond to realizations of lagged outcomes. However, many applications of difference-in-differences and event studies feature \emph{aggregate} covariates such as state-level policy variables. It seems unlikely that an aggregate treatment, chosen by, say, a local government, will respond to idiosyncratic shocks to individual outcomes. Does this imply that feedback bias is less of an issue in such settings?  

To examine this question, consider the following model with an aggregate treatment,
\begin{equation}
	Y_{it}=X_{j(i),t}\beta+A_{i}+F_{t}+U_{it},\quad i=1,...,N,\quad   t=1,...,T,\label{eq_aggreg}
\end{equation}
where $j(i)\in\{1,...,J\}$ is, say, the state where $i$ lives, and the treatment $X_{j(i),t}$ is determined at the state level (such as a minimum wage policy). Suppose that 
$$U_{it}=\zeta V_{j(i),t}+\varepsilon_{it},$$
where $V_{j(i),t}$ are shocks common to all individuals in state $j(i)$ at time $t$, $\varepsilon_{it}$ are purely idiosyncratic shocks, i.i.d. over $i$ and $t$, and $\zeta$ is a constant that measures the relative importance of the two components. 

It may be realistic to assume that \emph{idiosyncratic shocks} $\varepsilon_{it}$ are mean independent of the aggregate treatment $X_{j(i),s}$ in all periods $s=1,...,T$, i.e.,
$$\mathbb{E}[\varepsilon_{it}\,|\, {X_{j(i),1}},...,X_{j(i),T}]=0.$$
Hence, in a world without common shocks (i.e., $\zeta=0$), strict exogeneity of an aggregate treatment appears \textit{a priori} plausible, and standard difference-in-differences techniques may be justified.

However, \emph{common shocks} are present in many applications, and it is often restrictive to rule out all types of dynamic dependence of the treatment on them. To see this, suppose that there are many individual observations within a state, and write state-level averages based on (\ref{eq_aggreg}):
\begin{equation}
	\overline{Y}_{jt}=X_{jt}\beta+\overline{A}_{j}+F_{t}+\zeta V_{jt},\quad j=1,...,J,\quad t=1,...,T,\label{eq_aggreg_2}
\end{equation}
where $\overline{Y}_{jt}$ is the population average outcome in state $j$ at time $t$, and $\overline{A}_{j}$ is the population average of $A_i$ in state $j$. Note that the state-level average of the idiosyncratic shocks $\varepsilon_{it}$ is equal to zero. In (\ref{eq_aggreg_2}), both treatment and outcome are at the same level of aggregation (the state). Hence, the usual concerns with strict exogeneity assumptions apply. 

While it may be unrealistic to think of the decision of a local government to be based on individual outcome realizations $Y_{i,t-s}$, it is natural to expect that average realizations in the state $\overline{Y}_{j,t-s}$ (for example, how state-level employment evolved in the past) may influence how the state sets the minimum wage at time $t$. Now, if sequential exogeneity holds
 at the state level,  $$\mathbb{E}[V_{jt}\,|\, X_{j1},...,X_{jt}]=0,$$
but strict exogeneity does not hold,
  $$\mathbb{E}[V_{jt}\,|\, X_{j1},...,X_{jT}]\neq 0,$$
  and if $\zeta\neq 0$, then standard difference-in-differences estimators such as two-way fixed-effects will be biased and inconsistent in general. 
  
  This discussion highlights that concerns with strict exogeneity apply equally to aggregate treatments, unless the researcher can convincingly rule out the presence of dynamic selection based on aggregate shocks.

\section{Feedback in two- and three-period models\label{sec_2periods}}

The two-period model with a binary treatment is often presented as the canonical difference-in-differences setting (\citealp{abadie2005semiparametric}). Here we first analyze how the presence of feedback modifies the usual conclusions based on this model, and then discuss how to interpret a lack of pre-trends in a setting with an additional pre-treatment period.

\subsection{Bias in a two-period model with a binary treatment}

Suppose $X_{it}$ is binary, $X_{i1} = 0$ for all $i$, and $X_{i2} = 1$ for a subset of units (the treated group) while the other units (the control group) have $X_{i2} = 0$. Suppose that 
$$Y_{it} = B_{it}X_{it}+A_i+F_t+U_{it},\quad i=1,...,N,\quad t=1,2,$$
where note that the treatment effect $B_{it}$ is heterogeneous across units and over time. Lastly, suppose that the treatment is sequentially exogenous (but may not be strictly exogenous), so that
$$\mathbb{E}[U_{i2}\,|\, X_{i1}=0,X_{i2}=1]=\mathbb{E}[U_{i2}\,|\, X_{i1}=0,X_{i2}=0].$$

In this model, we can write
\begin{align*}
	&\overset{\text{DID}}{\overbrace{\mathbb{E}[Y_{i2}-Y_{i1}\,|\, X_{i1}=0,X_{i2}=1]-\mathbb{E}[Y_{i2}-Y_{i1}\,|\, X_{i1}=0,X_{i2}=0]}}\\&=\overset{\text{ATT}}{\overbrace{\mathbb{E}[B_{i2}\,|\, X_{i1}=0,X_{i2}=1]}}\\
	&\quad +\underset{\text{bias}}{\underbrace{\mathbb{E}[U_{i1}\,|\, X_{i1}=0,X_{i2}=0]-\mathbb{E}[U_{i1}\,|\, X_{i1}=0,X_{i2}=1]}},
\end{align*}
where the bias term is only zero under strict exogeneity. Indeed, in this setting, SE coincides with the assumption of parallel trends (PT), since, denoting as $Y_{it}(0)$ the potential outcome in the absence of treatment {(which, in the model above, depends on the treatment in period $t$ only)}, we have
\begin{align}
&\mathbb{E}[Y_{i2}(0)-Y_{i1}(0)\,|\, X_{i1}=0,X_{i2}=1]-\mathbb{E}[Y_{i2}(0)-Y_{i1}(0)\,|\, X_{i1}=0,X_{i2}=0]\notag\\&=	\mathbb{E}[U_{i1}\,|\, X_{i1}=0,X_{i2}=0]-\mathbb{E}[U_{i1}\,|\, X_{i1}=0,X_{i2}=1]=\text{bias}.\label{eq_PT}
	\end{align}

Under SE/PT, the bias term is equal to zero, and the difference-in-differences (DID) estimand is equal to the average treatment effect on the treated (ATT). This recovers the classic result in \citet{abadie2005semiparametric}. However, when SE/PT fails the bias term is generally non-zero, and the difference-in-differences estimand differs from the ATT in general.\footnote{Exceptions include the case where $U_{it}$ follows a random walk with an innovation that is independent of the treatment, as discussed in \citet{ghanem2022selection}.} 

\subsection{Pre-trends in a three-period model with a binary treatment}

Next, maintaining the setup from the previous subsection, suppose there is an additional initial period ``$0$'', where no-one is treated. It is common in applications to verify that pre-trends are parallel in support of the SE/PT assumption, as discussed in Subsection \ref{subsec_preT}. In the present model, pre-trends are parallel if  
\begin{align}
	&\mathbb{E}[Y_{i1}(0)-Y_{i0}(0)\,|\, X_{i1}=0,X_{i2}=1]-\mathbb{E}[Y_{i1}(0)-Y_{i0}(0)\,|\, X_{i1}=0,X_{i2}=0]\notag\\&=	{\mathbb{E}[U_{i1}-U_{i0}\,|\, X_{i1}=0,X_{i2}=1]-\mathbb{E}[U_{i1}-U_{i0}\,|\, X_{i1}=0,X_{i2}=0]=0.}\label{eq_preT}
\end{align}
Does the absence of pre-trends in (\ref{eq_preT}) imply that the SE/PT assumption holds, i.e., that the bias is zero in (\ref{eq_PT})? 

To illustrate the difference between parallel \emph{pre}-trends and parallel trends/SE, suppose $X_{i0}=X_{i1}$ is exogenously set to zero in periods $0$ and $1$, and treatment in period $2$ satisfies
$$X_{i2}=\boldsymbol{1}\left\{Z_{i}\geq 0\right\},$$
where selection into treatment depends on an index $Z_{i}$ that may include past shocks to outcomes $U_{i0}$ and $U_{i1}$, and some other unrelated factors, but not the current shocks $U_{i2}$ (consistently with the assumption of sequential exogeneity). In addition, suppose that the conditional means $\mathbb{E}[U_{i1}\,|\, Z_{i}]=\lambda_1(Z_{i}-\mathbb{E}[Z_i])$ and $\mathbb{E}[U_{i0}\,|\, Z_{i}]=\lambda_0(Z_{i}-\mathbb{E}[Z_i])$ are linear.  

In this case, pre-trends are parallel if and only if $\lambda_1=\lambda_0$ (by (\ref{eq_preT})), while the bias due to failure of SE/PT is (by (\ref{eq_PT})): 
\begin{align*}
\text{bias}	&=	\mathbb{E}[U_{i1}\,|\, X_{i1}=0,X_{i2}=0]-\mathbb{E}[U_{i1}\,|\, X_{i1}=0,X_{i2}=1]\\
&=\mathbb{E}[U_{i1}\,|\, Z_i<0]-\mathbb{E}[U_{i1}\,|\, Z_i\geq 0]\\
&=\mathbb{E}[\mathbb{E}\left(U_{i1}\,|\, Z_i\right)\,|\, Z_i<0]-\mathbb{E}[\mathbb{E}\left(U_{i1}\,|\, Z_i\right)\,|\, Z_i\geq 0]\\
&=\mathbb{E}[\lambda_1(Z_i-\mathbb{E}[Z_i])\,|\, Z_i<0]-\mathbb{E}[\lambda_1(Z_i-\mathbb{E}[Z_i])\,|\, Z_i\geq 0],
	\end{align*}
that is,
\begin{equation}
	\text{bias}	=\lambda_1\left(\mathbb{E}[Z_i\,|\, Z_i< 0]-\mathbb{E}[Z_i\,|\, Z_i\geq 0]\right).\label{eq_PT2}
\end{equation}

There are many instances where pre-trends are parallel, yet SE/PT does not hold and the bias in (\ref{eq_PT2}) is substantial. A simple example is when $Z_i=U_{i0}+U_{i1}+\xi_i$, where $U_{i0}$ and $U_{i1}$ have the same variances and are independent of $\xi_i$ (and conditional means are linear). In fact, it is clear in this example that there is no reason at all for $\lambda_1=\lambda_0$ to imply that the bias in (\ref{eq_PT2}) is zero. {The intuition is that, in periods $0$ and $1$, the treatment is exogenously set to zero, so there is no feedback to detect. The pre-trend test only asks whether selection into period-$2$ treatment is related to the \emph{change} $U_{i1}-U_{i0}$, whereas, under sequential exogeneity, the bias in (\ref{eq_PT}) depends on how selection relates to the \emph{level} $U_{i1}$. These two need not be related in general.}

%\begin{equation}
%	\limfunc{Cov}\left(U_{i1}-U_{i0},a_0U_{i0}+a_1U_{i1}\right)=0,\label{eq_preT_2}
%\end{equation}
%while the bias due to failure of SE/PT is, by (\ref{eq_PT}),
%\begin{equation}
%\text{bias}	=\limfunc{Cov}\left(U_{i1},a_0U_{i0}+a_1U_{i1}\right).\label{eq_PT2}
%\end{equation}

The discussion in this section underscores that failure of strict exogeneity causes bias and inconsistency even in the canonical two-period difference-in-differences design. \citet{ghanem2022selection} and \citet{marx2024parallel} provide in-depth analyses of selection issues in DID settings. Moreover, the simple three-period model considered here highlights that verifying that pre-trends are parallel does not suffice to support the assumption of strict exogeneity/parallel trends. 

The remainder of this article presents methods that, unlike two-way fixed effects estimators and related methods, are robust to failure of strict exogeneity and the presence of feedback. These methods rely on using lagged covariates as instrumental variables. Such methods are ineffective in the simple two- and three-period models with an absorbing treatment considered in this section since $X_{i0}$ and $X_{i1}$ do not vary at all. However, in less stylized settings where $X_{it}$ exhibits more variation, these methods provide ways to meaningfully relax strict exogeneity by allowing for feedback.

\section{Solutions and pitfalls in linear panel data models with constant coefficients\label{sec_AB}}

\subsection{Sequential moment restrictions}

A classic approach to estimation in linear models with sequentially exogenous covariates is based on \emph{sequential moment restrictions}. \citet{arellano2003panel} provides a comprehensive review of many available approaches. Consider the panel data model (\ref{PD_model}) under the assumption that covariates $X_{it}$ are sequentially exogenous as in (\ref{eq_seq}). In applications, $X_{it}$ may include lagged outcomes in addition to other sequentially exogenous covariates.\footnote{Including outcome lags is often crucial in applications, as recently highlighted in \citet{klosin2024dynamic}.} 

Differencing between periods $t$ and $t-1$ yields
\begin{equation}\mathbb{E}[Y_{it}-Y_{i,t-1}-(X_{it}-X_{i,t-1})'\beta\,|\, X_{i}^{t-1}]=0,\label{eq_seq_moment}\end{equation}
where in the rest of the article the notation $Z^t$ will indicate the history of $Z_t$ up to period $t$. Equation (\ref{eq_seq_moment}) represents a set of sequential conditional moment restrictions, for all periods $t$. As a result, any function $g(X_i^{t-1})$ of the history of the covariate can be used as an instrument in first differences, giving the unconditional moment restrictions
\begin{equation}\mathbb{E}[g(X_{i}^{t-1})\left(Y_{it}-Y_{i,t-1}-(X_{it}-X_{i,t-1})'\beta\right)]=0.\label{eq_seq_moment_uncond}\end{equation}

Different choices of instrument functions are used in practice. One example is the use of first differences as instruments, such as $g(X_{i}^{t-1})=X_{i,t-1}-X_{i,t-2}$. Another example is the use of instruments in levels, such as $g(X_{i}^{t-1})=X_{i,t-1}$. The popular GMM estimator from \citet{arellano1991some} (ABond hereafter) is based on the moment restrictions (\ref{eq_seq_moment}); see also \citet{holtz1988estimating}. The optimally-weighted GMM estimator achieves the semi-parametric efficiency bound of the model under restrictions (\ref{eq_seq_moment}) (see \citealp{chamberlain1992comment}).

%Can be applied to models with lagged outcomes. 
%$$\mathbb{E}[Y_{it}-Y_{i,t-1}-(X_{it}-X_{i,t-1})'\beta-\rho(Y_{i,t-1}-Y_{i,t-2})\,|\, X_{i}^{t-1},Y_i^{t-2}]=0.$$
%

%Can use IV in differences.
%
%$$\mathbb{E}[((X_{i,t-1}-X_{i,t-2}))(Y_{it}-Y_{i,t-1}-(X_{it}-X_{i,t-1})'\beta)]=0.$$
%
%
%
%Can use IV in levels.
%
%
%$$\mathbb{E}[X_{i,t-1}(Y_{it}-Y_{i,t-1}-(X_{it}-X_{i,t-1})'\beta)]=0.$$

\subsection{Issues with common dynamic panel data estimators}

Despite their popularity in applied work, estimators based on first differences and instrumental variables suffer from several shortcomings. Two main issues have been extensively discussed in the literature; see Chapter 11 in \citet{wooldridge2010econometric} for example. 

The first issue is that instruments may be \emph{weak}. For example, with $T=2$ and a scalar covariate we have
$$Y_{i2}-Y_{i1}=\beta(X_{i2}-X_{i1})+U_{i2}-U_{i1},$$
and the instrument $X_{i1}$ is weak when $X_{i2}-X_{i1}$ and $X_{i1}$ are weakly correlated. As another example, consider a first-order autoregressive model 
\begin{equation}Y_{it}=\rho Y_{i,t-1}+(1-\rho)A_i+U_{it},\label{eq_autoreg}\end{equation}
where $\mathbb{E}[U_{it}\,|\, Y_{i}^{t-1}]=0$. The relevance of {$Y_{i,t-2}$} as an instrument hinges on there being a non-zero correlation between $Y_{i,t-1}-Y_{i,t-2}$ and $Y_{i,t-2}$. One thus expects the instrument to be weak when $\rho$ is close to 1.

A second issue with panel data estimators based on IV and first differences is that, when $T$ is not small relative to $N$, the number of instruments is \emph{large} relative to the sample size. Indeed, the fact that any lagged $X_{it}$ beyond the first order is a valid instrument because of (\ref{eq_seq_moment}) implies that the number of instruments increases with $T$. Specifically, linear moment restrictions imply $T(T-1)/2$ valid moments (for a single scalar covariate). This causes an issue of proliferation of instruments. In simulations, researchers have found that method-of-moment estimators exhibit bias in such cases (e.g., \citealp{kiviet1995bias}). 

Considering the autoregressive model (\ref{eq_autoreg}) in an asymptotic regime where both $N$ and $T$ increase, \citet{alvarez2003time} show that the ABond estimator is biased when $N/T$ tends to a constant. Specifically, they show that, while the OLS estimator with fixed effects suffers from a bias of order 1/T, the ABond estimator suffers from a bias of order 1/N. When $N$ and $T$ have comparable magnitudes, the bias is not negligible relative to the standard deviation of the estimator (which is proportional to $1/\sqrt{NT}$). This distorts confidence intervals and affects the validity of inference. 

The analysis in \citet{alvarez2003time} implies that, when $N$ and $T$ have comparable magnitudes, IV estimators may suffer from biases, similarly to OLS with fixed effects. This issue is due to the fact that the estimator is based on a set of moment restrictions that grows with $T$. If instead one uses a single lag $Y_{i,t-2}$ in model (\ref{eq_autoreg}), for example, then the GMM estimator does not suffer from an asymptotic bias in the asymptotic regime analyzed in \citet{alvarez2003time}.

\subsection{Additional moment restrictions}

A first reaction to the weak instruments issue highlighted in the previous subsection has been to impose additional moment restrictions. One popular assumption is that the association between the instruments and the unobserved effects $A_i$ is constant. For example, when using lags of $X_{it}$ as instruments this requires assuming that  
\begin{equation}\mathbb{E}[X_{it}A_i]=\mathbb{E}[X_{is}A_i] \text{ for all }t,s.\label{eq_stat}\end{equation}
If the stationarity condition (\ref{eq_stat}) holds, then $$\mathbb{E}[(X_{it}-X_{i,t-1})(A_i+U_{it})]=\underset{=0\text{ by }(\ref{eq_stat})}{\underbrace{\mathbb{E}[X_{it}A_i]-\mathbb{E}[X_{i,t-1}A_i]}}+\mathbb{E}[(X_{it}-X_{i,t-1})\underset{=0\text{ by } (\ref{eq_seq})}{\underbrace{\mathbb{E}[U_{it}\,|\, X_i^t]}}] =0,$$
so $X_{it}-X_{i,t-1}$ is a valid instrument for the equations in levels, leading to the moment conditions
\begin{equation}
	\mathbb{E}[(X_{it}-X_{i,t-1})(Y_{it}-X_{it}'\beta)]=0.\label{eq_stat_2}
\end{equation}

\citet{arellano1995another} introduced such stationarity restrictions and analyzed efficiency properties in a setting that combines restrictions of the form (\ref{eq_stat_2}) with the ABond restrictions (\ref{eq_seq_moment}), corresponding to the ``system GMM'' estimator. \citet{blundell1998initial} adapted and applied this strategy for production function estimation.

Another type of moment restrictions was introduced by \citet{ahn1995efficient}. Suppose that the sequential exogeneity assumption (\ref{seqpan1}) is replaced by
\begin{equation}\mathbb{E}[U_{it}\,|\, X_i^t,Y_{i}^{t-1},A_i]=0.\label{eq_AS}\end{equation}
This requires adding lagged outcomes as regressors to the model, which is often not restrictive since dynamic models typically include lagged outcomes. In addition, (\ref{eq_AS}) requires mean independence to hold conditional on the unobserved effect $A_i$. While this \emph{conditional} sequential exogeneity assumption is formally more restrictive than its unconditional version $\mathbb{E}[U_{it}\,|\, X_i^t,Y_{i}^{t-1}]=0$, it still allows for unrestricted forms of feedback. Now, if (\ref{eq_AS}) holds, then
$$\mathbb{E}[(A_i+U_{it})(U_{i,t-1}-U_{i,t-2})]=0,$$
which implies quadratic moment restrictions on the parameters. Such restrictions may be combined with the ones derived by \citet{arellano1991some} and \citet{arellano1995another}. 

However, while one may hope that adding informative moment restrictions may alleviate weak instruments issues, by construction doing so \emph{increases} the number of moment restrictions used in estimation and may thus be subject to the ``many instruments'' problem. The methods reviewed next are not based on additional moment restrictions and are thus potentially more robust to this last issue.  

\subsection{Quasi-likelihood approaches}

Another reaction to the issues with the ABond and other GMM estimators based on sequential moment restrictions has been to develop alternative estimation strategies. A strategy proposed by several authors is to adopt a likelihood approach based on a Gaussian specification. The motivation for this approach is that the likelihood implicitly ``weighs'' the sequential moments implied by the model in a way that would be optimal under the Gaussian parametric assumptions, yet may perform better than GMM even when the parametric assumptions are violated (hence the name ``quasi'' likelihood). 

 \citet{hsiao2002maximum} introduce a transformed maximum likelihood approach for equations in first differences. \citet{alvarez2003time} consider and analyze a random-effects estimation approach in levels that additionally requires a normal specification for the distribution of the individual effect $A_i$ given initial conditions; see also \citet{alvarez2022robust}. This correlated random-effects specification is reminiscent of the \citet{mundlak1978pooling} approach for static panel data models. \citet{moral2013likelihood} and \citet{moral2019dynamic} develop likelihood approaches for general models with sequentially exogenous regressors. \citet{bun2017maximum} study the finite-sample performance of some of these estimators in simulations.

Consider the autoregressive model in \citet{alvarez2003time},
\begin{equation}Y_{it}=\rho Y_{i,t-1}+A_i+U_{it},\quad i=1,...,N,\quad t=1,...,T.\label{eq_autoreg2}\end{equation}
Suppose that $A_i$, given the initial condition $Y_{i0}$, is drawn from a Gaussian distribution with a mean that is linear in $Y_{i0}$ and a constant variance, and suppose that $U_{it}$ are Gaussian i.i.d. over time with constant variance. Maximizing the log-likelihood yields the maximum likelihood estimator $\widehat{\rho}$, which is consistent and asymptotically normal as $N$ tends to infinity under Gaussianity. Moreover, $\widehat{\rho}$ remains consistent and asymptotically normal even if the true data generating process does not satisfy Gaussian assumptions, albeit with a different asymptotic variance.\footnote{\citet{alvarez2022robust} consider the case where the variance of $U_{it}$ varies over time.} \citet{alvarez2003time} additionally show that, as $N$ and $T$ tend to infinity at the same rate,
$$\sqrt{NT}(\widehat{\rho}-\rho)\overset{d}{\rightarrow}{\cal{N}}(0,1-\rho^2),$$
which implies that, unlike the ABond estimator, the quasi-maximum likelihood estimator $\widehat{\rho}$ does not suffer from asymptotic bias.

In more general models with sequentially exogenous covariates, \citet{moral2013likelihood} proposes a likelihood approach that relies on a specification of the feedback process, assuming that
$$X_{it}\,|\, Y_{i}^{t-1},X_{i}^{t-1},A_i$$
is Gaussian with a mean that depends linearly on $Y_{i}^{t-1},X_{i}^{t-1},A_i$ and a constant variance matrix. As in the case of the likelihood specification for the autoregressive model, the resulting quasi-maximum likelihood estimator remains consistent and asymptotically normal even when the Gaussian assumptions are violated.

\subsection{Large-T perspective and bias correction\label{subsec_largeT}}

The large-T perspective introduced in \citet{hahn2002asymptotically} can be used to develop novel estimators with improved properties in panels of moderate size (i.e., where the time dimension is not too short). This class of estimators is based on an asymptotic approach where $N$ and $T$ tend to infinity jointly. The key idea is, starting from an estimator such as OLS, to then remove the bias (or, in practice, an estimate of the bias) from the estimator. 

To illustrate the idea, consider the simple case of the OLS estimator with fixed effects $\widehat{\rho}$ based on the autoregressive model (\ref{eq_autoreg2}). As $N,T$ tend to infinity at the same rate, \citet{hahn2002asymptotically} show that 
\begin{equation}
	\sqrt{NT}\left(\widehat{\rho}-\left(\rho-\frac{1+\rho}{T}\right)\right)\overset{d}{\rightarrow}{\cal{N}}\left(0,1-\rho^2\right).
\end{equation}
This suggests that the bias-corrected estimator
$$\widehat{\rho}^{\rm BC}=\widehat{\rho}+\frac{1+\widehat{\rho}}{T}$$
satisfies
\begin{equation}
	\sqrt{NT}\left(\widehat{\rho}^{\rm BC}-\rho\right)\overset{d}{\rightarrow}{\cal{N}}\left(0,1-\rho^2\right).\label{as_dist_rho_hat}
\end{equation}
In other words, the bias-corrected estimator $\widehat{\rho}^{\rm BC}$ is free from (asymptotic) bias, and the asymptotic distribution of $\sqrt{NT}\left(\widehat{\rho}^{\rm BC}-\rho\right)$ is correctly centered at zero, enabling the usual construction of asymptotically valid confidence intervals.

More generally, in models with general sequentially exogenous covariates, and starting with an estimator $\widehat{\beta}$ satisfying an expansion of the form (\ref{bias_expand}), the bias-correction approach is based on a consistent estimator $\widehat{C}$ of the constant $C$, and on the resulting bias-corrected estimator
$$\widehat{\beta}^{\rm BC}=\widehat{\beta}-\frac{\widehat{C}}{T}.$$
Typically, the asymptotic distribution of $\sqrt{NT}(\widehat{\beta}^{\rm BC}-\beta)$ is correctly centered, similar to (\ref{as_dist_rho_hat}). See \citet{arellano2007understanding} for a survey of various approaches, and \citet{dhaene2015split} for an approach based on split-panel jackknife. 

The price to pay for relying on large-T asymptotic arguments is that bias-corrected estimators such as $\widehat{\rho}^{\rm BC}$ and $\widehat{\beta}^{\rm BC}$ are generally not consistent in short panels, in the sense that they do not converge to the population parameter as $N$ tends to infinity while $T$ is fixed, only as $N$ and $T$ tend to infinity jointly. In some settings, it is possible to completely remove the bias and achieve fixed-$T$ consistency, as \citet{bun2005bias} do in an autoregressive model, however this situation appears to be the exception rather than the rule.

\subsection{Misspecification issues in dynamic panel models}

Except for methods based on large-T arguments, all the methods reviewed in this section rely on conditional moment restrictions such as (\ref{eq_seq_moment}) being satisfied. Hence, the credibility of the methods hinges on the validity of instrumental variables that, unlike in the quasi-experimental literature, are \emph{internal} to the model. As a result, a threat to validity is that the model be misspecified in some dimensions. One form of misspecification, which was already highlighted in \citet{arellano1991some}, is related to the dependence structure of errors when lagged outcomes are present in the equation. Indeed, considering the autoregressive model (\ref{eq_autoreg2}) and assuming that $\mathbb{E}[U_{it}\,|\, Y_i^{t-1}]=0$ requires $U_{it}$ to be independent over time. \citet{arellano1991some} propose a test of this hypothesis, against a more general alternative that $U_{it}$ follow a moving-average process of a given order. 

Empirically, there are other reasons why the model, and the implied moment restrictions, may be misspecified. One reason is that, unlike what model (\ref{PD_model}) postulates, the coefficients $\beta$ may be heterogeneous across individuals and possibly over time as well (\citealp{marx2024heterogeneous}). This concern with treatment effects \emph{heterogeneity} has taken a central place in modern applied econometrics, as reviewed in the introduction, and it is the topic of the next section. Another possible source of misspecification is that, in contrast with the linear model (\ref{PD_model}), the true relationship between outcomes $Y_{it}$ and covariates $X_{it}$ may be \emph{nonlinear}. Such a concern is prevalent in settings with continuous treatments or discrete outcomes, and it is the topic of Section \ref{sec_nonlin}.

\section{Heterogeneity and feedback\label{sec_hetero}}

\subsection{Chamberlain's negative result}

Allowing for coefficient heterogeneity is of paramount importance in many applications. In policy evaluation settings, how to handle the presence of unobserved treatment effects heterogeneity is often a central part of the analysis. In models with strictly exogenous covariates, many methods have been developed to estimate features of the distribution of heterogeneous coefficients; see, e.g., \citet{chamberlain1992efficiency}, \citet{arellano2012identifying}, \citet{graham2012identification}, and the survey by \citet{bonhomme2025fixed}. Those methods have also been applied to difference-in-differences settings; see \citet{borusyak2024revisiting}, \citet{botosaru2025time}, \citet{de2025treatment}, and the survey by \citet{arkhangelsky2024causal}.

However, identification raises important new challenges when covariates are sequentially exogenous and treatment effects are heterogeneous. An important contribution by Gary Chamberlain (\citealp{chamberlain2022feedback}) establishes a negative identification result in a model with a binary sequentially exogenous covariate where both the coefficient of the covariate and the intercept vary unrestrictedly across individuals.

To understand the identification challenge, consider the following random coefficients model
\begin{equation}
	Y_{it}=B_iX_{it}+A_i+F_t+U_{it},\quad i=1,...,N,\quad t=1,...,T,\label{eq_random_slope}
\end{equation}
where it is assumed that the scalar covariate $X_{it}$ is sequentially exogenous in the sense of
\begin{equation}\mathbb{E}\left[U_{it}\,|\, X_{i1},...,X_{it}\right]=0,\label{eq_random_slope_ass}\end{equation}
and observations are i.i.d. over $i$. Suppose one wishes to characterize the identified set for the average treatment effect parameter $\mu=\mathbb{E}[B_i]$. Establishing identification requires showing that the set is in fact a singleton. 

To provide intuition, consider first model (\ref{eq_random_slope}) without time effects (i.e., $F_t=0$) and $T=2$, and write first differences as
\begin{equation}
	Y_{i2}-Y_{i1}=B_i(X_{i2}-X_{i1})+U_{i2}-U_{i1}.
\end{equation}
Clearly, the mean of $B_i$ may only be identified in subpopulations such that $X_{i2}-X_{i1}$ is non-zero. In other words, the average treatment effect (ATE) is never identified, and as is common in the literature we will aim to identify some form of average treatment effect ``on the treated'' (ATT). In the present discussion I will assume that the support of $X_{i2}-X_{i1}$ is bounded away from zero.\footnote{\citet{graham2012identification} study the case where $X_{i2}-X_{i1}$ is close to zero and identification may be irregular.}

We then have
\begin{equation}
	\underset{=\widehat{B}_i}{\underbrace{\frac{Y_{i2}-Y_{i1}}{X_{i2}-X_{i1}}}}=B_i+\frac{U_{i2}-U_{i1}}{X_{i2}-X_{i1}},
\end{equation}
where $\widehat{B}_i$ is the OLS estimator of $B_i$ in model (\ref{eq_random_slope}) for the \emph{single} individual $i$. Since there are only two periods, we expect $\widehat{B}_i$ to be very noisy. Nevertheless, under strict exogeneity, the average OLS (or ``mean group'') estimator is unbiased for $\mu$, since
$$\mathbb{E}\left[\frac{1}{N}\sum_{i=1}^N  \widehat{B}_i\right]=\mathbb{E}[B_i]+\frac{1}{N}\sum_{i=1}^N\mathbb{E}\left[\frac{\overset{=0}{\overbrace{\mathbb{E}[U_{i2}-U_{i1}\,|\, X_{i1},X_{i2}]}}}{X_{i2}-X_{i1}}\right]=\mu.$$
However, this argument breaks down when $X_{it}$ are not strictly exogenous and $$\mathbb{E}[U_{i2}-U_{i1}\,|\, X_{i1},X_{i2}]\neq 0.$$ 

\citet{chamberlain2022feedback} considers model (\ref{eq_random_slope}) with a binary treatment in the presence of time effects, and he takes $T=2$. That is, imposing the normalization $F_1=0$ and denoting $F_2=F$,
\begin{align}
	Y_{i1}&=B_iX_{i1}+A_i+U_{i1},\quad \mathbb{E}[U_{i1}\,|\, X_{i1}]=0,\label{eq_chamb1}\\
	Y_{i2}&=B_iX_{i2}+A_i+F+U_{i2},\quad \mathbb{E}[U_{i2}\,|\, X_{i1},X_{i2}]=0.\label{eq_chamb2}
\end{align}
Obviously, in this model there is no way to identify $\mathbb{E}[B_i\,|\, X_{i1}=1,X_{i2}=1]$ and $\mathbb{E}[B_i\,|\, X_{i1}=0,X_{i2}=0]$ (``average treatment effects on stayers'') due to the presence of the unrestricted intercept $A_i$. However, remarkably, \citet{chamberlain2022feedback} shows that $\mathbb{E}[B_i\,|\, X_{i1}=0,X_{i2}=1]$ and $\mathbb{E}[B_i\,|\, X_{i1}=1,X_{i2}=0]$ (``average treatment effects on movers'') are not identified either, nor is the time effect $F$. This situation stands in sharp contrast with settings under strict exogeneity, where these last three quantities are identified, although average effects on stayers are not (\citealp{chamberlain1992efficiency}).

\subsection{Point-identified averages}

There have been two responses to Chamberlain's negative result. The first one has been to characterize and study quantities that remain identified under sequential {exogeneity}. This is the topic of this subsection and the next. The second response, which I will turn to in Subsection \ref{subsec_partial}, has been to adopt a partial identification approach and bound the quantities that are not point-identified.

Consider model (\ref{eq_random_slope})-(\ref{eq_random_slope_ass}), with the normalization $F_T=0$, and focus for concreteness on the case where $X_{it}$ is binary. As in \citet{bonhomme2025unrestricted}, we ask what weighted averages of $B_i$,
$$\mu=\mathbb{E}\left[c(X_{i1},...,X_{iT})B_i\right],$$
are identified, where $c(X_{i1},...,X_{iT})$ are possibly positive or negative (scalar) weights.

A sufficient condition for $\mu$ to be identified is that 
there exist functions $\varphi_t:\{0,1\}^t\mapsto \mathbb{R}$, for $t=1,...,T-1$, such that
\begin{align}
	c(X_{i1},...,X_{iT})&=\sum_{t=1}^{T-1}(X_{it}-X_{iT})\varphi_t(X_{i}^t),\label{prop_eq1}\\
	0&=\mathbb{E}[\varphi_t(X_{i}^t)],\quad t=1,...,T-1.\label{prop_eq2}
\end{align}
Here, $\varphi_t(X_{i}^t)$ are functions of the history of the covariate, which are effectively used as instrumental variables (as is often the case in dynamic panel data settings). {Future values $X_{i,t+s}$, for $s\geq 1$, cannot be used when strict exogeneity is violated.}

To see why (\ref{prop_eq1}) and (\ref{prop_eq2}) imply that $\mu$ is identified, note that
\begin{align*}
	\mu&=\mathbb{E}\left[c(X_{i1},...,X_{iT})B_i\right]\\
	&=\mathbb{E}\left[\left(\sum_{t=1}^{T-1}(X_{it}-X_{iT})\varphi_t(X_{i}^t)\right)B_i\right] \text{ by }(\ref{prop_eq1})\\
	&=\mathbb{E}\left[\sum_{t=1}^{T-1}\varphi_t(X_{i}^t)(X_{it}B_i-X_{iT}B_i)\right] \\
	&=\mathbb{E}\left[\sum_{t=1}^{T-1}\varphi_t(X_{i}^t)(Y_{it}-F_{t}-U_{it}-Y_{iT}+U_{iT})\right]\\
	&=\mathbb{E}\left[\sum_{t=1}^{T-1}\varphi_t(X_{i}^t)(Y_{it}-Y_{iT})\right]-\sum_{t=1}^{T-1}\mathbb{E}\left[\varphi_t(X_{i}^t)\right]F_{t}\text{ by }(\ref{eq_random_slope_ass})\\
	&=\mathbb{E}\left[\sum_{t=1}^{T-1}\varphi_t(X_{i}^t)(Y_{it}-Y_{iT})\right]\text{ by }(\ref{prop_eq2}),		
\end{align*}
where the next-to-last line uses that, by (\ref{eq_random_slope_ass}) and the law of iterated expectations,
$$\mathbb{E}\left[\varphi_t(X_{i}^t)(U_{it}-U_{iT})\right]=\mathbb{E}\left[\varphi_t(X_{i}^t)\mathbb{E}\left(U_{it}-U_{iT}\,|\, X_i^t\right)\right]=0,$$
and the expression in the last line is a population mean of observed variables, hence identified. Note that (\ref{prop_eq2}) is only needed due to the presence of the time effects $F_t$.

In fact, it can be shown using the strategy in \citet{bonhomme2025unrestricted} that, if no functions $\varphi_t$ satisfy (\ref{prop_eq1})-(\ref{prop_eq2}), then $\mu$ is not identified. Moreover, in that case the identified set for $\mu$ is the whole real line. Hence, the class of weights $c$ leading to point-identified weighted averages is fully characterized as the solution to a linear system of equations whose parameters are the values of the functions $\varphi_t$. This allows for simple and exhaustive characterizations in cases that have been previously considered in the literature.

{As a first example, consider model (\ref{eq_chamb1})-(\ref{eq_chamb2}) without time effects and $T=2$. In this model, (\ref{prop_eq2}) is not needed for identification (due to the absence of time effects), and (\ref{prop_eq1}) reads
\begin{align}
	c(X_{i1},X_{i2})&=(X_{i1}-X_{i2})\varphi_1(X_{i1}).
\end{align}
This implies that both average treatment effects on movers are identified -- that is, for $(X_{i1},X_{i2})=(1,0)$ and for $(0,1)$ -- although the average effects on stayers are not identified, exactly as in the strictly exogenous case. Identification can be seen from the fact that
$$\mathbb{E}\left[(Y_{i1}-Y_{i2})\varphi_1(X_{i1})\right]=\mathbb{E}\left[(X_{i1}-X_{i2})\varphi_1(X_{i1})B_i\right]+\underset{=0}{\underbrace{\mathbb{E}\left[(U_{i1}-U_{i2})\varphi_1(X_{i1})\right]}}. $$

As a second example, consider the same model now including time effects, as in \citet{chamberlain2022feedback}. In this case (\ref{prop_eq1}) and (\ref{prop_eq2}) read, denoting $\pi_1=\mathbb{E}[X_{i1}]$, 
\begin{align*}
	c(X_{i1},X_{i2}) &= X_{i1}(X_{i1}-X_{i2})\varphi_1(1)
	+ (1-X_{i1})(X_{i1}-X_{i2})\varphi_1(0),\\
	0 &= \pi_1\varphi_1(1) + (1-\pi_1)\varphi_1(0).
\end{align*}
Hence, {if $\pi_1\in(0,1)$,} $\mu$ is point-identified if and only if $c(X_{i1},X_{i2}) $ is proportional to \begin{equation}{c}_0(X_{i1},X_{i2})=(1-\pi_1)X_{i1}(1-X_{i2})+{\pi_1}(1-X_{i1})X_{i2}.\label{eq_czero}\end{equation}
This defines a one-dimensional linear space that includes neither the average effects on movers nor the ones on stayers, consistently with the analysis in \citet{chamberlain2022feedback}.}

{As a third example, consider again the model without time effects but now with $T=3$, as studied in \citet{arellano2012identifying}. In this model, (\ref{prop_eq1}) reads}
\begin{align}
	c(X_{i1},X_{i2},X_{i3})&=(X_{i1}-X_{i3})\varphi_1(X_{i1})+(X_{i2}-X_{i3})\varphi_2(X_{i1},X_{i2}).
\end{align}
Hence, enumerating the $2^3=8$ support points of $(X_{i1},X_{i2},X_{i3})$, the identification condition can equivalently be written as
\begin{align}
	c(X_{i1},X_{i2},X_{i3})&=\varphi_1(1)X_{i1}(X_{i1}-X_{i3})+\varphi_1(0)(1-X_{i1})(X_{i1}-X_{i3})+\varphi_2(1,1)X_{i1}X_{i2}(X_{i2}-X_{i3})\notag\\
	&+\varphi_2(1,0)X_{i1}(1-X_{i2})(X_{i2}-X_{i3})+\varphi_2(0,1)(1-X_{i1})X_{i2}(X_{i2}-X_{i3})\notag\\
	&+\varphi_2(0,0)(1-X_{i1})(1-X_{i2})(X_{i2}-X_{i3}),\label{eq_c_AB}
\end{align}
which defines a six-dimensional linear space. {\citet{arellano2012identifying} highlight that $X_{i1}(X_{i2}-X_{i1})$, $X_{i1}(X_{i3}-X_{i1})$, and $X_{i2}(X_{i3}-X_{i2})$ all lead to point identification. In fact, as one can easily see from (\ref{eq_c_AB}), all average treatment effects on movers -- that is, excluding $(0,0,0)$ and $(1,1,1)$ -- are identified. However, the scope for identification is severely reduced when time effects are present in the model.}

\subsection{Best identified approximation}

When point-identification of $\mu$ fails, its identified set is unbounded. In this case, following \citet{azriel2020estimation}, \citet{bonhomme2025unrestricted} proposes to focus on the weighted average $\mu^*$ that is closest to $\mu$ while being point-identified. The quantity $\mu^*$ is the \emph{best identified approximation} to $\mu$.

To illustrate this approach, consider again model (\ref{eq_chamb1})-(\ref{eq_chamb2}), and suppose that the researcher is interested in the average treatment effect $\mu=\mathbb{E}[B_i]$, which corresponds to $c$ being a vector of ones. Although $\mu$ is not point-identified {in} that setting, one can consider 
$$\mu^*=\mathbb{E}\left[\lambda^* {c}_0(X_{i1},X_{i2})B_i\right],$$
where $c_0$ is given by (\ref{eq_czero}), and 
$$\lambda^*=\underset{\lambda}{\limfunc{argmin}}\, \mathbb{E}\left[\left(1-\lambda{c}_0(X_{i1},X_{i2})\right)^2\right].$$
 Note the presence of the constant weights $c=1$ in this formula, which correspond to the average effect of interest $\mathbb{E}[B_i]=\mathbb{E}[1\times B_i]$. 
  
\begin{figure}[h!]
	\caption{Revisiting Chamberlain's negative result: best identified weights}\label{fig_chamb}
\begin{center}
	\includegraphics[width=10cm]{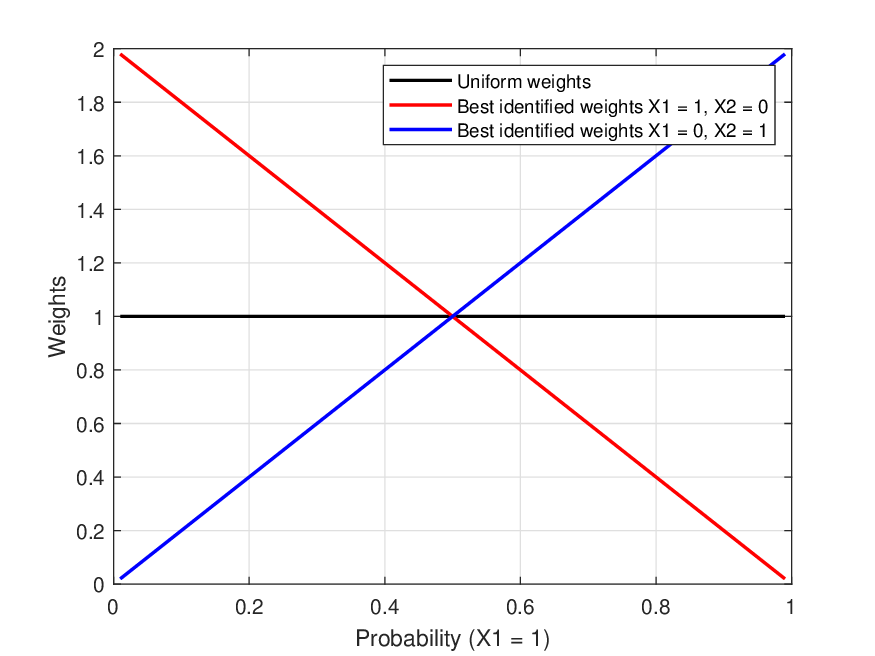}
\end{center}
{\footnotesize{\textit{Notes: Uniform weights $c$ normalized to 1 in black, best identified weights $c^*$ in the subpopulation with $X_{i1}=0,X_{i2}=1$ in blue, best identified weights $c^*$ in the subpopulation with $X_{i1}=1,X_{i2}=0$ in red. $X_{i2}\sim\mbox{Ber}(0.5)$ independent of $X_{i1}$.}}} 
\end{figure}

I plot the weights $c=1$ and {$c^*=\lambda^*c_0$} in Figure \ref{fig_chamb}, for the case where $X_{i1}$ and $X_{i2}$ follow independent Bernoulli distributions: $X_{i1}\sim \mbox{Ber}(\pi_1)$, where I vary $\pi_1$ on the x-axis, and $X_{i2}\sim \mbox{Ber}(1/2)$. The original weights $c$ are identically equal to one, and are shown in black in the Figure. The weights $c^*$ corresponding to the best identified approximation are shown in blue (for $c^*(0,1)$) and red (for $c^*(1,0)$). The best identified weights for stayers (not shown in the Figure) are $c^*(0,0)=c^*(1,1)=0$.

Several conclusions can be drawn from Figure \ref{fig_chamb}. First, $c^*$ and $c$ are different, which reflects the lack of point-identification of $\mu=\mathbb{E}[B_i]$. Second, despite these differences, the weights $c^*$ corresponding to the best identified approximation are all non-negative, consistent with a ``weakly causal'' interpretation of $\mu^*$. Third, as the probability $\pi_1=\Pr(X_{i1}=1)$ increases, $\mu^*$ puts more weight on the combination $(X_{i1}=0,X_{i2}=1)$ relative to $(X_{i1}=1,X_{i2}=0)$. Lastly, when $\pi_1=\frac{1}{2}$ we have $c^*(0,1)=c^*(1,0)=1$, which identifies the following average treatment effect on movers: $$\mathbb{E}[B_i\,|\, (X_{i1}=0,X_{i2}=1) \text{ OR }(X_{i1}=1,X_{i2}=0)].$$   

Characterizing further the properties of best identified approximations in models with sequential exogeneity is a promising avenue for future work. 

\subsection{Partial identification: the (Wooyong) Lee bounds\label{subsec_partial}}

To see how to obtain finite bounds on $\mu$ even when it is not point-identified, consider again model (\ref{eq_random_slope}) -- without time effects for simplicity -- and assume that sequential exogeneity holds conditional on the latent individual effects, 
\begin{equation}\mathbb{E}\left[U_{it}\,|\, X_{i1},...,X_{it},A_i,B_i\right]=0.\label{eq_seq_cond}\end{equation}
Note that this conditional sequential exogeneity assumption differs from the unconditional one, (\ref{eq_random_slope_ass}), which we assumed in the previous two subsections.

When (\ref{eq_seq_cond}) holds, the characterization of the identified set is more complicated than in (\ref{prop_eq1})-(\ref{prop_eq2}). The reason is that one needs to work with the entire distributions of the latent variables $(A_i,B_i)$, as their conditional expectations no longer exhaust all the information in the model. Nevertheless, one can show that point identification fails in many cases. For example, focusing on an autoregressive model with $X_{it}=Y_{i,t-1}$, \citet{lee2026identification} shows that $\mathbb{E}[B_i]$ is not point-identified. Hence, imposing the stronger condition (\ref{eq_seq_cond}) is not enough  to restore point-identification.\footnote{{\citet{pesaran2024heterogeneous} show how to consistently estimate moments of the heterogeneous coefficient in a first-order autoregressive model under the additional assumption that initial conditions are stationary.}} 

Nevertheless, under (\ref{eq_seq_cond}) one can bound the average coefficient $\mathbb{E}[B_i]$, as shown by \citet{lee2026identification}. To see why, consider the simple case $T=2$, and note the model implies the following unconditional moment restrictions
\begin{equation}\mathbb{E}[(A_i+B_iX_{it})(Y_{it}-B_iX_{it}-A_i)]=0,\quad t=1,2.\label{eq_m_wooyong}\end{equation}
It turns out that (\ref{eq_m_wooyong}) is enough to obtain finite bounds on $\mu=\mathbb{E}[B_i]$.

To develop the argument, observe that, for all scalar $\lambda$,
$$
\mu=\mathbb{E}\left[B_i+\lambda\sum_t(A_i+B_iX_{it})(Y_{it}-B_iX_{it}-A_i)\right].
$$
Hence, denoting
$$
Q_i(A_i,B_i,\lambda)=B_i+\lambda\sum_t(A_i+B_iX_{it})(Y_{it}-B_iX_{it}-A_i),
$$
we have
$$
\mu=\mathbb{E}[Q_i(A_i,B_i,\lambda)].
$$

Next, note that $Q_i(A_i,B_i,\lambda)$ is a second-order polynomial in $A_i,B_i$. Taking $\lambda>0$, we have
$$
\underset{A_i}{\limfunc{max}}\, Q_i(A_i,B_i,\lambda)
= B_i+\lambda\sum_t\!\left(\tfrac{1}{2}\overline{Y}_i+B_i(X_{it}-\overline{X}_i)\right)
\!\left(Y_{it}-\tfrac{1}{2}\overline{Y}_i-B_i(X_{it}-\overline{X}_i)\right)=P_i(B_i,\lambda),
$$
where $P_i(B_i,\lambda)$ is a second-order polynomial in $B_i$. The coefficient of $B_i^2$ in $P_i(B_i,\lambda)$ is $-\lambda\sum_t(X_{it}-\overline{X}_i)^2$. If $\sum_t(X_{it}-\overline{X}_i)^2>0$ then the polynomial is strictly concave and it has a finite maximum:
$$
Z_i(\lambda)=\underset{B_i}{\limfunc{max}}\, P_i(B_i,\lambda)=\underset{A_i,B_i}{\limfunc{max}}\, Q_i(A_i,B_i,\lambda).
$$
Hence, if $Z_i(\lambda)$ has finite mean then
\begin{equation}
	\mu=\mathbb{E}\left[Q_i(A_i,B_i,\lambda)\right]\leq \mathbb{E}\left[Z_i(\lambda)\right].
	\label{eq_bound}
\end{equation}
This provides a finite upper bound on $\mu=\mathbb{E}[B_i]$. The argument for the lower bound is similar, taking $\lambda<0$.

This basic bound can be improved in many ways. One approach is to optimize with respect to $\lambda$. Another approach is to consider additional moment restrictions implied by the model.\footnote{In fact, the model implies a continuum of those, since one can take arbitrary functions of $A_i$, $B_i$, and $X_i^t$ as instruments.} See \citet{lee2026identification} for a detailed analysis of identification and inference on the identified set. %See also \citet{chesher2024robust} and \citet{botosaru2024adversarial} for approaches based on partial identification in static and dynamic panel data models. 

The existence of a finite bound on $\mu$ contrasts with the discussion following (\ref{prop_eq1})-(\ref{prop_eq2}) that, under (\ref{eq_random_slope_ass}), the identified set of $\mu$ is either a singleton or the whole real line. The reason is that the bound in (\ref{eq_bound}) is based on (\ref{eq_seq_cond}), which imposes sequential exogeneity \emph{conditional} on $A_i,B_i$. Suppose instead that sequential exogeneity holds unconditionally, as in (\ref{eq_random_slope_ass}). In this case, the only moment restrictions implied from the model are of the form
$$
\mathbb{E}[\psi(X_i^t)(Y_{it}-B_iX_{it}-A_i)]=0,
$$
for arbitrary functions $\psi$. Then, defining analogously to before
$$
Q_i(A_i,B_i,\lambda)=B_i+\lambda\psi(X_i^t)(Y_{it}-B_iX_{it}-A_i),
$$
one notes that $Q_i(A_i,B_i,\lambda)$ is a \emph{first-order} polynomial in $A_i,B_i$ that has no finite maximum (except in trivial cases). Hence, indeed, $\mu$ is either point-identified or its identified set is the whole real line in this case.

\subsection{Restoring point-identification?}

The failure of point-identification in model (\ref{eq_random_slope}) arises since heterogeneity $(A_i,B_i)$ is bivariate. A possibility to restore point-identification is to assume that heterogeneity is \emph{scalar}. Consider as an example the model 
\begin{equation}
	Y_{it}=B_iX_{it}+A_i+U_{it},\quad i=1,...,N,\quad t=1,...,T,
\end{equation}
where one assumes that
$$B_i=\theta_1 A_i+\theta_0.$$
This implies
$$Y_{it}=\theta_0X_{it}+A_i(1+\theta_1X_{it})+U_{it}.$$

Hence we have (assuming that the denominator is non-zero)
$$\frac{Y_{it}-\theta_0X_{it}}{1+\theta_1X_{it}}=A_i+V_{it},$$
where $V_{it}=\frac{U_{it}}{1+\theta_1X_{it}}$ is such that
$\mathbb{E}[V_{it}\,|\, X_i^t]=0$. Taking first differences then implies the conditional moment restrictions
$$\mathbb{E}\left[\frac{Y_{it}-\theta_0X_{it}}{1+\theta_1X_{it}}-\frac{Y_{i,t-1}-\theta_0X_{i,t-1}}{1+\theta_1X_{i,t-1}}\,\bigg|\, X_i^{t-1}\right]=0.$$
Instruments in levels and differences can be used to identify and estimate $\theta_0$ and $\theta_1$. Once those have been recovered, one obtains
$$\mathbb{E}[A_i]=\mathbb{E}\left[\frac{Y_{it}-\theta_0X_{it}}{1+\theta_1X_{it}}\right],\quad \mathbb{E}[B_i]=\theta_1 \mathbb{E}[A_i]+\theta_0.$$
\citet{chamberlain2022feedback} characterizes efficient estimators in this setup. However, assuming that $(A_i,B_i)$ is uni-dimensional is a strong assumption, which restricts the nature of heterogeneity substantially.

Another possibility to guarantee point-identification in models with coefficient heterogeneity and sequential {exogeneity} is to follow a \emph{large-$T$} approach, as outlined in Subsection \ref{subsec_largeT}. In this approach, estimates based on models with coefficient heterogeneity and sequentially exogenous covariates typically satisfy bias expansions of the form (\ref{bias_expand}). Given this, bias correction approaches can be applied to such settings as well; see \citet{fernandez2013panel} for details. A drawback of large-T approaches is that they do not achieve fixed-T consistency in general, and thus may not perform well in short panels.

\section{Nonlinearity and feedback\label{sec_nonlin}}

Economic models often imply nonlinear relationships between covariates and outcomes. Decreasing returns, or curvature of preferences, generate nonlinearities that have implications for policy predictions. However, a challenge to extend the analysis in the previous sections to nonlinear settings comes from the presence of latent heterogeneity (i.e., ``fixed effects''). Unlike in linear models, it is typically not possible to difference out the heterogeneity except in special cases (\citealp{arellano2001panel}). Under strict exogeneity, solutions exist in general semi-parametric likelihood models (\citealp{bonhomme2012functional}). However, the presence of un-modeled dynamics and feedback creates additional challenges. 

After mentioning some model-specific approaches in certain nonlinear settings, this section focuses on semi-parametric likelihood models with feedback, reviewing results on the existence of valid moment restrictions (which are useful when the parameters are point-identified) and on the characterization of identified sets more generally. The last part of the section incorporates restrictions on the feedback process and shows how this affects the analysis.

\subsection{Moment conditions in some specific models}

Moment restrictions allowing for feedback have been derived in certain specific nonlinear models with a multiplicative structure. \citet{Blundell_Griffith_Windmeijer_JOE2002} consider count data regression models; see also \citet{wooldridge1997multiplicative}. \citet{al2017exponential} derive moment conditions in a binary choice model with an exponential structure. The efficiency analysis in \citet{chamberlain2022feedback} covers those cases. 

To see how to handle the presence of multiplicative heterogeneity and feedback, consider a panel data model with a multiplicative structure, 
$$Y_{it}=\exp(\beta X_{it}+A_i)U_{it},$$
where 
$$\mathbb{E}[U_{it}\,|\, X_i^t,A_i]=1.$$
Note that here we require a moment restriction that is \emph{conditional} on the individual heterogeneity $A_i$. Multiplying by $\exp(-\beta X_{it})$ gives
$$\mathbb{E}[\exp(-\beta X_{it})Y_{it}\,|\, X_i^t,A_i]=\exp(A_i).$$
Hence, taking differences between periods $t$ and $t-1$, we obtain
$$\mathbb{E}[\exp(-\beta X_{it})Y_{it}-\exp(-\beta X_{i,t-1})Y_{i,t-1}\,|\, X_i^{t-1},A_i]=\exp(A_i)-\exp(A_i)=0.$$

Generalizing this approach to models without a multiplicative structure has proven difficult. In unpublished dissertation work, \citet{woutersen2000essays} {derives} moment conditions in a multiple-spell mixed proportional hazards model of duration. Recently, \citet{bonhomme2023identification} study identification in binary choice models with feedback. They find that lack of point-identification is pervasive in these models. For example, studying a logit binary choice model
$$Y_{it}=\boldsymbol{1}\{\beta X_{it}+A_i+U_{it}\geq 0\},$$
where $X_{it}$ are binary and $$U_{it}\,|\, X_i^t,Y_{i}^{t-1},A_i\sim\text{Logistic},$$ they find that $\beta$ is never point-identified, irrespective of how large $T$ is. This finding contrasts with the case with a strictly exogenous $X_{it}$, where point-identification can be achieved when $T=2$ (\citealp{georg1960probabilistic}), and with the autoregressive case $X_{it}=Y_{i,t-1}$, where point-identification can be achieved when $T=3$ (\citealp{chamberlain2023identification}).

\subsection{An approach for likelihood models}

\citet{bonhomme2025feedback} introduce an approach to find moment conditions in nonlinear models with feedback that have a semi-parametric likelihood structure, such that the outcome follows a parametric distribution indexed by a parameter vector $\theta$,
 $$Y_{it}\,|\, Y_{i,t-1},X_{it},A_i\sim f_{\theta},$$ and the feedback process and the distribution of heterogeneity given initial conditions are both unrestricted. This setup mimics, in a nonlinear setting, linear dynamic models with sequential exogeneity where the model restricts the mean (and possibly the variance) of outcomes given covariates and heterogeneity while leaving feedback and heterogeneity unrestricted.

%
%
%\begin{align}
%	\ell\left(y^{T},x^{T};\theta,g,\pi,\nu\right)= &\bigg(\int  \prod_{t=1}^{T}\underset{\text{Parametric component}}{\underbrace{f_{\theta}\left(\left.y_{t}\right|y_{t-1},x_{t},a\right)}}\times\prod_{t=2}^{T}\underset{\text{Feedback process}}{\underbrace{g\left(\left.x_{t}\right|y^{t-1},x^{t-1},a\right)}}\nonumber \\
%	&\quad\quad \quad \times\underset{\text{Heterogeneity}}{\underbrace{\pi\left(a\, |\,y_{0},x_{1}\right)}}da\bigg)\times\underset{\text{Initial condition}}{\underbrace{\nu\left(y_{0},x_{1}\right)}},\label{eq: feedback_complete_data_likelihood}
%\end{align}
%where the density $f_{\theta}$ of the outcome is parametric, whereas the feedback process $g$, the heterogeneity density $\pi $, and the density $\nu$ of initial conditions are all unrestricted. 

\citet{bonhomme2025feedback} provide necessary and sufficient conditions for a moment function $\phi_{\theta}$ to be \emph{feedback and heterogeneity robust (FHR)}, in the sense that 
\begin{equation}\mathbb{E}[\phi_{\theta}(Y_{i}^T,X_i^T)]=0,\label{eq_phi}\end{equation}
irrespective of the form of the feedback process and the distribution of heterogeneity. Their characterization, which takes the form of integral equations, extends the characterization under strict exogeneity in \citet{bonhomme2012functional} to models with feedback. 

To provide intuition about the conditions ensuring the FHR property, consider first the case $T=2$ under the assumption that covariates are strictly exogenous. \citet{bonhomme2012functional} points out that $\phi_{\theta}$ satisfies (\ref{eq_phi}) if and only if
\begin{equation}\mathbb{E}[\phi_{\theta}(Y_{i0},Y_{i1},Y_{i2},X_{i1},X_{i2})\,|\,Y_{i0},X_{i1},X_{i2},A_i ]=0.\label{eq_phi_SE}\end{equation}
  That (\ref{eq_phi_SE}) implies (\ref{eq_phi}) is obvious by the law of iterated expectations, and the converse holds due to the fact that the distribution of $A_i$ given covariates is unrestricted. Now, note that (\ref{eq_phi_SE}) solely depends on $\theta$. Indeed, (\ref{eq_phi_SE}) is equivalent to 
\begin{equation}\iint  \phi_{\theta}(y_0,y_{1},y_{2},x_{1},x_{2})f_{\theta}(y_2\,|\, x_2,y_1,a)f_{\theta}(y_1\,|\, x_1,y_0,a)dy_2dy_1=0.\label{eq_phi_SE_int}\end{equation}
Hence, finding moment functions $\phi_{\theta}$ amounts to solving the integral equation (\ref{eq_phi_SE_int}). \citet{bonhomme2012functional}, and subsequently \citet{honore2024moment}, \citet{honore2025dynamic}, and \citet{dano2023transition}, among others, use this approach to find moment restrictions in a variety of models with strictly exogenous covariates. 

\citet{bonhomme2025feedback} observe that, in models with sequentially exogenous covariates, (\ref{eq_phi_SE_int}) is no longer sufficient for $\phi_{\theta}$ to provide a valid moment restriction in general. This is because, in the presence of feedback, $X_{i2}$ depends on $Y_{i1}$ and (\ref{eq_phi_SE_int}) is no longer equivalent to (\ref{eq_phi_SE}). The authors show that, for $\phi_{\theta}$ to be a valid moment function, it needs to satisfy another condition, which in integral form reads
\begin{equation}\int \phi_{\theta}(y_0,y_{1},y_{2},x_{1},x_{2})f_{\theta}(y_2\,|\, x_2,y_1,a)dy_2\text{ does not depend on }x_2.\label{eq_phi_SeqE_int}\end{equation}
They show that the FHR moment functions are those that satisfy both (\ref{eq_phi_SE_int}) and (\ref{eq_phi_SeqE_int}). The second condition (\ref{eq_phi_SeqE_int}) can be interpreted as ensuring robustness to the presence of an unknown feedback process. It can be equivalently stated as
\begin{equation*}\mathbb{E}[\phi_{\theta}(Y_{i0},Y_{i1},Y_{i2},X_{i1},X_{i2})\,|\,Y_{i0},Y_{i1},X_{i1},X_{i2},A_i ]=\mathbb{E}[\phi_{\theta}(Y_{i0},Y_{i1},Y_{i2},X_{i1},X_{i2})\,|\,Y_{i0},Y_{i1},X_{i1},A_i ],\end{equation*}
which requires that $X_{i2}$ does not predict $\phi_{\theta}$, conditional on its other arguments and the individual effect $A_i$.

\citet{bonhomme2025feedback} show that FHR moment functions span the orthogonal complement of the nuisance tangent set of the model, and are thus helpful to construct efficient moment functions using sequential projection arguments, as in the literature on linear dynamic models initiated by \citet{chamberlain1992comment} and \citet{arellano1995another}. In addition, they provide an analogous characterization of FHR moment functions for average effects of the form 
\begin{equation}\mu=\mathbb{E}[h_{\theta}(Y_{i}^T,X_i^T,A_i)],\label{eq_ave}\end{equation}
for a known function $h$, hence allowing one to obtain estimators of average partial effects and other average effects of interest in economic applications.

\subsection{Identified sets in nonlinear models}

In certain models, feedback and heterogeneity robust (FHR) moment functions as in (\ref{eq_phi}) may fail to exist. In such cases, a natural approach is to resort to partial identification and bound the quantity of interest. Fortunately, identified sets in likelihood models with feedback have a tractable representation. To see this, let $f_0(y^T,x^T)$ denote the joint density of $Y_t,X_t$ across periods. Focusing again on the case $T=2$ for simplicity, let 
$$g(x_2\,|\, y_1,y_0,x_1,a)$$
denote the feedback process (i.e., the conditional density of $X_{i2}$), let
$$\pi(a\,|\,y_0,x_1)$$
denote the density of heterogeneity $A_i$, 
and let
$$\nu(y_0,x_1)$$
denote the density of initial conditions $Y_{i0},X_{i1}$. Consider as before a setting where $g$, $\pi$, and $\nu$ are all left unrestricted. The joint likelihood of $(Y_{i2},Y_{i1},Y_{i0},X_{i2},X_{i1},A_i)$ is
\begin{align*}f_{\theta}(y_2\,|\, y_1,x_2,a)g(x_2\,|\, y_1,y_0,x_1,a)f_{\theta}(y_1\,|\, y_0,x_1,a)\pi(a\,|\, y_0,x_1)\nu(y_0,x_1),
	\end{align*}
and the identified set for $\theta$ is the set of $\theta$ values such that, for some $g,\pi,\nu$, 
\begin{equation}
	f_0(y_0,y_1,y_2,x_1,x_2)=\int f_{\theta}(y_2\,|\, y_1,x_2,a)g(x_2\,|\, y_1,y_0,x_1,a)f_{\theta}(y_1\,|\, y_0,x_1,a)\pi(a\,|\, y_0,x_1)\nu(y_0,x_1)
da.\label{eq_IS}\end{equation}
In words, the identified set is the set of $\theta$ values that are consistent with the data density for some values of the feedback process, the heterogeneity density, and the density of initial conditions. 

\citet{bonhomme2023identification} observe that (\ref{eq_IS}) is equivalent to the existence of a density $p(y_0,y_1,y_2,x_1,x_2,a)$ such that the following conditions hold:
\begin{align}
	f_0(y_0,y_1,y_2,x_1,x_2)&=\int p(y_0,y_1,y_2,x_1,x_2,a)da,\label{eq_IS1}\\
	 p(y_0,y_1,y_2,x_1,x_2,a)&=f_{\theta}(y_2\,|\, y_1,x_2,a)\int  p(y_0,y_1,y_2,x_1,x_2,a)dy_2,\label{eq_IS2}\\
	 \iint p(y_0,y_1,y_2,x_1,x_2,a)dy_2dx_2&=f_{\theta}(y_1\,|\, y_0,x_1,a) \iiint p(y_0,y_1,y_2,x_1,x_2,a)dy_2dy_1dx_2,\label{eq_IS3}
\end{align}
 where (\ref{eq_IS1}) expresses that the model and data density are consistent with each other, while (\ref{eq_IS2}) and (\ref{eq_IS3}) express that the model admits $f_{\theta}$ as the (parametric) conditional densities of outcomes in both periods.

Since (\ref{eq_IS1}),  (\ref{eq_IS2}) and (\ref{eq_IS3}) are linear in the density $p$, one can check whether any given value $\theta$ belongs to the identified set by verifying whether a linear program has a solution. This feature was first noticed by \citet{honore2006bounds} in likelihood models with strictly exogenous covariates, and it is here extended to models with feedback. Likewise, \citet{bonhomme2023identification} show that, for $\theta$ fixed, the identified set of an average effect such as (\ref{eq_ave}) can be computed by relying on linear programming. 

\begin{figure}[h!]
\caption{Identified sets in binary choice models with a binary covariate}\label{fig_sets}
\begin{center}
	\begin{tabular}{cc}
		\multicolumn{2}{c}{Logit}\\
		$T=2$ & $T=4$\\
		\includegraphics[width=50mm, height=50mm]{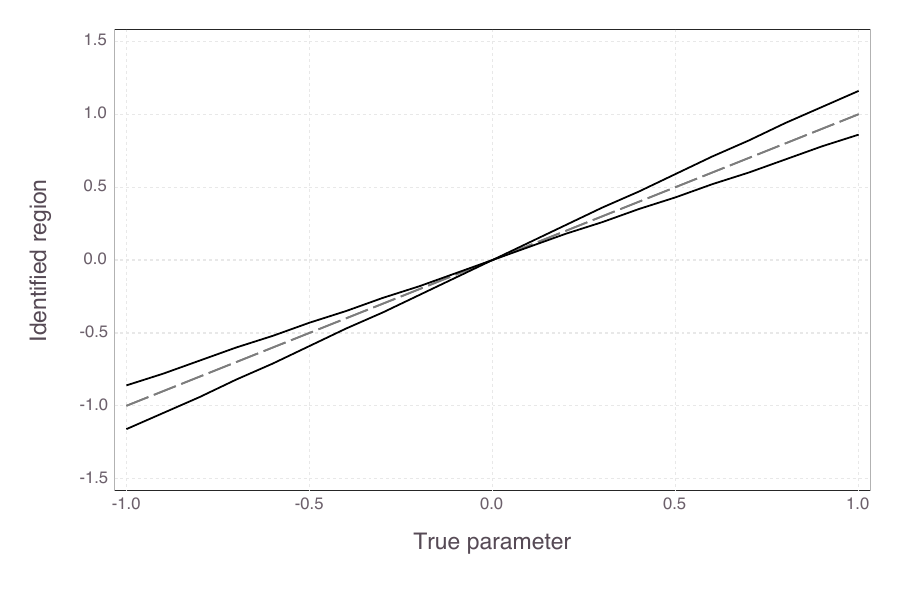} &	\includegraphics[width=50mm, height=50mm]{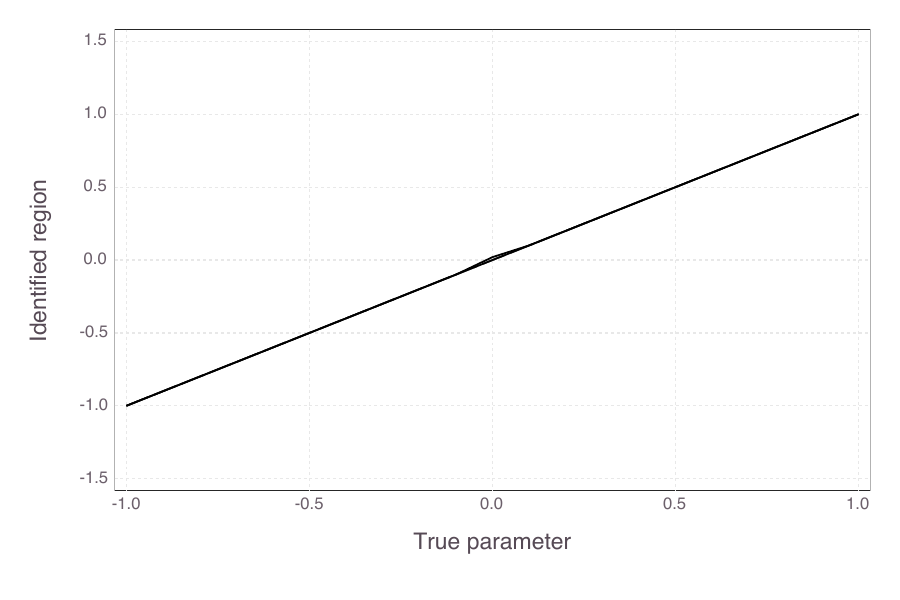}\\
	\multicolumn{2}{c}{Probit}\\
	$T=2$ & $T=4$\\
\includegraphics[width=50mm, height=50mm]{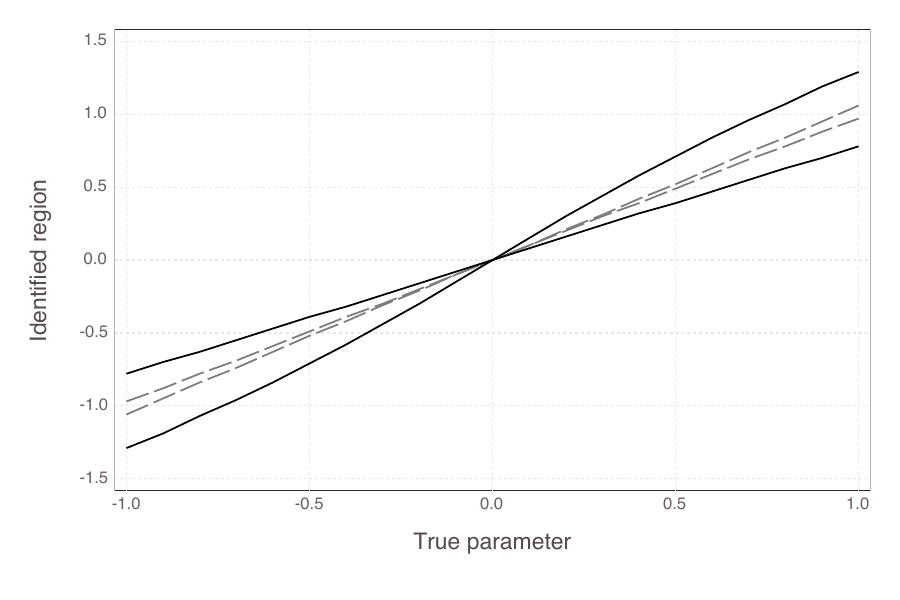} &	\includegraphics[width=50mm, height=50mm]{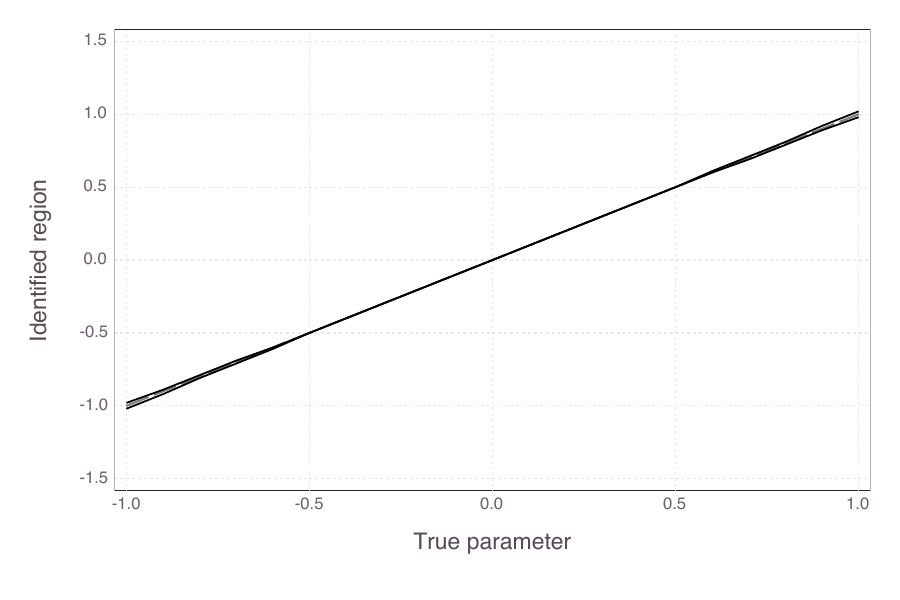}
\end{tabular}
\end{center}
\par
\textit{{\footnotesize Notes: From \citet{bonhomme2023identification}. The identified sets under strict exogeneity are indicated in dashed lines, the sets under sequential exogeneity are indicated in solid lines. The upper panel shows results for the logit model, the lower panel shows results for the probit model. $T=2$ is shown in the left column, $T=4$ in the right column.}}
\end{figure}

Figure \ref{fig_sets}, reproduced from \citet{bonhomme2023identification}, shows the identified sets for $\theta$ in the binary choice model 
$$Y_{it}=\boldsymbol{1}\{\theta X_{it}+A_i+U_{it}\geq 0\},$$
where $X_{it}$ are sequentially exogenous. The Figure was obtained under a particular data generating process, for which the true $\theta$ parameter is indicated on the x-axis. In the logit case (top panel), one sees that the identified set is a singleton under strict exogeneity (dashed line), yet that it is an interval with non-empty interior under sequential exogeneity (solid line). Moreover, going from $T=2$ to $T=4$ shrinks the size of the identified set substantially, to the point that it is essentially a singleton and $\theta$ is close to being point-identified. This observation suggests that, despite the failure of point-identification, identified sets may be informative in applications to (not too short) panels. In the probit case (bottom panel), the Figure shows similar conclusions, with the difference that point-identification fails under both strict and sequential exogeneity.  

Partial identification is a promising approach in nonlinear panel data models with strictly or sequentially exogenous covariates (with or without a likelihood structure). See \citet{botosaru2024adversarial} and \citet{chesher2024robust} for recent proposals.

\subsection{Restricted feedback}

Restrictions on the feedback process can have identifying power. Consider first the case of \emph{homogeneous feedback},
\begin{equation}X_{it}\text{ independent of }A_i\text{ given } Y_{i}^{t-1},X_{i}^{t-1},\label{eq_homog}\end{equation}
under which the feedback process does not depend on $A_i$ and is identical across individuals. \citet{robins1986new}, and subsequent work, develop approaches based on such sequential exchangeability assumptions.\footnote{{Recent work extends $g$-estimation methods developed under sequential exchangeability to settings where that assumption is replaced by parallel trends conditions allowing for unobserved time-invariant confounders (\citealp{shahn2022structural}, \citealp{renson2023identifying}).}} Condition (\ref{eq_homog}) is plausible in dynamic experiments, where the researcher controls the treatment $X_{it}$ and can adjust it depending on past outcome realizations. In non-experimental settings, it is \textit{a priori} less plausible, although it may be a natural assumption in some economic settings.\footnote{An example is when $X_{it}$ is a dynamic state variable (e.g., wealth) whose law of motion does not vary across individuals (e.g., if returns to wealth do not vary across individuals).}

%
%To see why homogeneous feedback (\ref{eq_homog}) gives rise to additional moment conditions, assume that
%\begin{equation}\mathbb{E}[U_{it}\,|\, X_i^{t},Y_i^{t-1},A_i,B_i]=0.\end{equation}
%This implies 
%$$\mathbb{E}[Y_{it}\,|\, X_i^t,Y_i^{t-1},A_i,B_i]=A_i+B_i X_{it},$$
%which are the only moment restrictions implied by sequential exogeneity. However, by homogeneous feedback (\ref{eq_homog}) we also have
%\begin{align*}
%	\mathbb{E}[Y_{it}\,|\, X_i^{t-1},Y_i^{t-1},A_i,B_i]&=A_i+B_i\mathbb{E}[X_{it}\,|\, X_i^{t-1},Y_i^{t-1},A_i,B_i]\\&=A_i+B_i\mathbb{E}[X_{it}\,|\, X_i^{t-1},Y_i^{t-1}],
%\end{align*} 
%which provides additional sets of moment conditions. 

Consider again the case $T=2$ for illustration. Under homogeneous feedback, the feedback process is
$$g(x_2\,|\, y_1,y_0,x_1),$$
independent of $a$. Hence, the characterization of the identified set for $\theta$ becomes (assuming that the denominator on the left-hand side is non-zero)
\begin{equation}
	\frac{f_0(y_0,y_1,y_2,x_1,x_2)}{g(x_2\,|\, y_1,y_0,x_1)\nu(y_0,x_1)}=\int f_{\theta}(y_2\,|\, y_1,x_2,a)f_{\theta}(y_1\,|\, y_0,x_1,a)\pi(a\,|\, y_0,x_1)da
	.\label{eq_IS_homog}\end{equation}
As \citet{bonhomme2023identification} point out, the right-hand side in (\ref{eq_IS_homog}) coincides with the likelihood of a model with strictly exogenous covariates (and $A_i$ independent of $X_{i2}$ conditional on $Y_{i0},X_{i1}$). This shows that, once the density of the data is properly weighted, one can use methods for the analysis of models with strictly exogenous covariates for identification and estimation.

Another possible restriction on the feedback process is \emph{Markovian feedback}, such as the first-order Markov restriction
	$$X_{it}\text{ independent of }Y_i^{t-2},X_i^{t-2}\text{ given } Y_{i,t-1},X_{i,t-1},A_i.$$
This type of restriction is common in dynamic structural models, and it has identifying content too. \citet{arellano2016nonlinear} study identification under such Markovian assumptions (building on previous work by \citealp{hu2008instrumental} and \citealp{hu2012nonparametric}), and propose an estimation approach based on quantile regressions.

\section{Looking ahead: feedback in networks}

Most of the panel data literature focuses on single-agent models where individuals do not interact with each other. However, in many empirical applications, interactions and spillovers are of economic interest. The concept of feedback is similarly relevant to network settings.

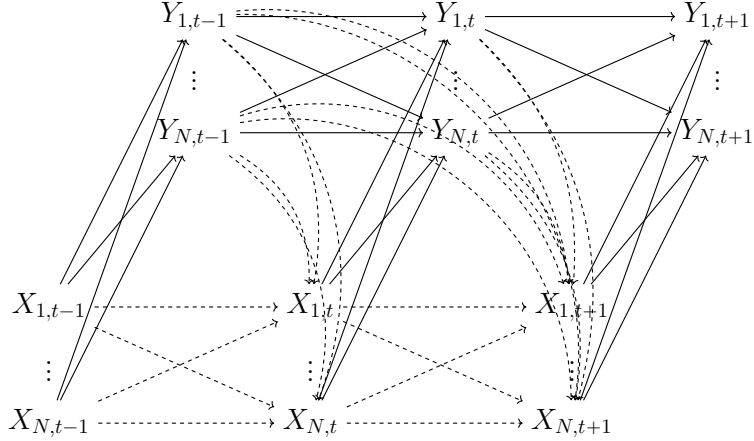
\begin{figure}
\begin{center}
	
		\caption{An illustration of sequential exogeneity in networks\label{fig_net}}

\resizebox{0.6\textwidth}{!}{%	
\begin{tikzpicture}[
	node distance=2cm and 5cm,
	every node/.style={font=\large},
	scale=1.5,
	transform shape
	]
	
	% Define horizontal offsets
	\def\xshiftX{0}
	\def\xshiftY{2.5}
	\def\dx{4.5}
	
	% Time t-1 (Moderate spacing)
	\node (x1tm1) at (\xshiftX, 2) {$X_{1,t-1}$};
	\node at (\xshiftX, 1) {$\vdots$};
	\node (xNtm1) at (\xshiftX, 0) {$X_{N,t-1}$};
	
	\node (y1tm1) at (\xshiftX+\xshiftY, 7) {$Y_{1,t-1}$};
	\node at (\xshiftX+\xshiftY, 6) {$\vdots$};
	\node (yNtm1) at (\xshiftX+\xshiftY, 5) {$Y_{N,t-1}$};
	
	% Time t (Moderate spacing)
	\node (x1t) at (\xshiftX+\dx, 2) {$X_{1,t}$};
	\node at (\xshiftX+\dx, 1) {$\vdots$};
	\node (xNt) at (\xshiftX+\dx, 0) {$X_{N,t}$};
	
	\node (y1t) at (\xshiftX+\xshiftY+\dx, 7) {$Y_{1,t}$};
	\node at (\xshiftX+\xshiftY+\dx, 6) {$\vdots$};
	\node (yNt) at (\xshiftX+\xshiftY+\dx, 5) {$Y_{N,t}$};
	
	% Time t+1 (Moderate spacing)
	\node (x1tp1) at (\xshiftX+2*\dx, 2) {$X_{1,t+1}$};
	\node at (\xshiftX+2*\dx, 1) {$\vdots$};
	\node (xNtp1) at (\xshiftX+2*\dx, 0) {$X_{N,t+1}$};
	
	\node (y1tp1) at (\xshiftX+\xshiftY+2*\dx, 7) {$Y_{1,t+1}$};
	\node at (\xshiftX+\xshiftY+2*\dx, 6) {$\vdots$};
	\node (yNtp1) at (\xshiftX+\xshiftY+2*\dx, 5) {$Y_{N,t+1}$};
	
	% Y to Y (self and cross)
	\draw[->, thick] (y1tm1) -- (y1t);
	\draw[->, thick] (y1tm1) -- (yNt);
	\draw[->, thick] (yNtm1) -- (y1t);
	\draw[->, thick] (yNtm1) -- (yNt);
	
	\draw[->, thick] (y1t) -- (y1tp1);
	\draw[->, thick] (y1t) -- (yNtp1);
	\draw[->, thick] (yNt) -- (y1tp1);
	\draw[->, thick] (yNt) -- (yNtp1);
	
	% X to X (dashed, self and cross)
	\draw[->, dashed, thick] (x1tm1) -- (x1t);
	\draw[->, dashed, thick] (x1tm1) -- (xNt);
	\draw[->, dashed, thick] (xNtm1) -- (x1t);
	\draw[->, dashed, thick] (xNtm1) -- (xNt);
	
	\draw[->, dashed, thick] (x1t) -- (x1tp1);
	\draw[->, dashed, thick] (x1t) -- (xNtp1);
	\draw[->, dashed, thick] (xNt) -- (x1tp1);
	\draw[->, dashed, thick] (xNt) -- (xNtp1);
	
	% Solid: X --> Y (within time)
	\foreach \src/\dst in {
		x1tm1/y1tm1, x1tm1/yNtm1, xNtm1/y1tm1, xNtm1/yNtm1,
		x1t/y1t, x1t/yNt, xNt/y1t, xNt/yNt,
		x1tp1/y1tp1, x1tp1/yNtp1, xNtp1/y1tp1, xNtp1/yNtp1}
	\draw[->, thick] (\src) -- (\dst);
	
	% Dashed: Y_{t-1} --> X_{t}
	\draw[->, dashed, thick] (y1tm1) to[bend left=30] (x1t);
	\draw[->, dashed, thick] (y1tm1) to[bend left=35] (xNt);
	\draw[->, dashed, thick] (yNtm1) to[bend left=30] (x1t);
	\draw[->, dashed, thick] (yNtm1) to[bend left=35] (xNt);
	
	% Dashed: Y_{t} --> X_{t+1}
	\draw[->, dashed, thick] (y1t) to[bend left=25] (x1tp1);
	\draw[->, dashed, thick] (y1t) to[bend left=30] (xNtp1);
	\draw[->, dashed, thick] (yNt) to[bend left=25] (x1tp1);
	\draw[->, dashed, thick] (yNt) to[bend left=30] (xNtp1);
	
	\draw[->, dashed, thick] (y1tm1) to[bend left=45] (x1tp1);
	\draw[->, dashed, thick] (y1tm1) to[bend left=50] (xNtp1);
	\draw[->, dashed, thick] (yNtm1) to[bend left=45] (x1tp1);
	\draw[->, dashed, thick] (yNtm1) to[bend left=50] (xNtp1);
	
\end{tikzpicture}
}

\end{center}

	{\footnotesize{\textit{Notes: Schematic description of sequential exogeneity in a network setting.}}}

\end{figure}

Let $\boldsymbol{Y}_t=(Y_{1t},...,Y_{Nt})$ and $\boldsymbol{X}_t=(X_{1t},...,X_{Nt})$. Let also $\boldsymbol{A}=(A_1,...,A_N)$. The feedback process is the density of 
$$ \boldsymbol{X}_t\,|\, \boldsymbol{Y}^{t-1},{\boldsymbol{X}^{t-1}},\boldsymbol{A}.$$
It is appealing to allow for general forms of feedback. For example, changes in network structure (i.e., $\boldsymbol{X}_t$) could be partly driven by shocks to outcomes (\citealp{kuersteiner2020dynamic}). An illustration of feedback in networks is given by Figure \ref{fig_net}. Models with unrestricted feedback allow past outcomes of all units, $Y_{i,t-1}$, to affect future unit-specific covariates $X_{i',t}$ (such as network links).

As an illustration, consider a model on a {bipartite network}, 
\begin{equation}\label{eq_AKM}
	Y_{it}=\sum_{j=1}^JB_j X_{ijt}+A_i+U_{it}, \quad i=1,...,N,\quad t=1,...,T,
\end{equation}
under the sequential exogeneity assumption
\begin{equation}\label{eq_AKM_ass} \mathbb{E}\left[U_{it}\,|\, X^t\right]=0,
\end{equation}
where $X^t$ denotes the set of all $X_{ijs}$ for $i=1,...,N$, $j=1,...,J$, and $s\leq t$. 

For example, $i$ and $j$ may denote workers and firms, respectively, as in the AKM model of wage determination (\citealp{abowd1999high}), and $X_{ijt}$ denote the indicator that $i$ works in $j$ at $t$. In this model, $A_i$ and $B_j$ are worker and firm effects, respectively. However, model (\ref{eq_AKM})-(\ref{eq_AKM_ass}) allows for feedback, since it does not impose the strict exogeneity assumption of the AKM model that stipulates \begin{equation}\mathbb{E}\left[U_{it}\,|\, X^T\right]=0.\label{eq_strictex}\end{equation}
While the so-called ``exogenous mobility'' condition (\ref{eq_strictex}) rules out job mobility to be influenced by previous wage shocks $U_{it}$, (\ref{eq_AKM_ass}) allows for such dependence.\footnote{Model (\ref{eq_AKM})-(\ref{eq_AKM_ass}) also differs from a ``distributed lags'' specification that allows past firms to affect future wages, yet still relies on strict exogeneity (\citealp{di2023ain}).}

\citet{bonhomme2019distributional} show that allowing for feedback is essential for the model to be compatible with {structural models} of sorting and wage determination. Here, the feedback process represents a dynamic model of {network} {formation} with heterogeneous workers and firms. An important advantage of leaving the feedback process unrestricted is that it is not necessary to specify a dynamic network formation model.

Estimation raises a number of issues, in part related to the existence of many moment restrictions. See \citet{mikusheva2025estimation}, \citet{kuersteiner2020dynamic}, and also \citet{bonhomme2019distributional} for a setting that allows for nonlinear relationships between wages and (discrete) worker and firm heterogeneity. The literature on feedback in networks is still in its infancy, and more work is needed given the potential relevance of these methods for economic applications.

\section{Concluding remarks}

Many popular estimation methods rely on (explicit or implicit) {restrictive assumptions} about feedback. Feedback is central to many economic models, and often empirically plausible. Yet, many popular methods hinge on the assumption of strict exogeneity that rules out feedback entirely. It is important for applied researchers to be {aware of the dynamic} {restrictions} they impose, since those are often key for identification.

Allowing simultaneously for feedback and heterogeneity raises difficult challenges. While the recent literature has made some progress on these and other vexing issues, feedback should become (once again) a {central topic} for econometric research.  

\clearpage

{\small
	\bibliographystyle{econometrica}
	\bibliography{biblio}
}

\end{document}